\documentclass[preprint,singlecolumn]{elsarticle}       
\usepackage{setspace}
\usepackage{bm}
\usepackage{graphicx}
\usepackage{natbib}
\usepackage[normalem]{ulem}
\usepackage{amsmath}
\usepackage{amssymb}
\usepackage{fullpage}
\usepackage{color}

\newcommand{\be}{\begin{equation}}
\newcommand{\ee}{\end{equation}}
\newcommand{\bea}{\begin{eqnarray}}
\newcommand{\eea}{\end{eqnarray}}

\bibpunct{(}{)}{;}{a}{}{,}

\begin{document}
\begin{frontmatter}
\title{Computational Modeling of Channelrhodopsin-2 Photocurrent Characteristics in Relation to Neural Signaling}
\author{Roxana A. ~Stefanescu\fnref{fn1}}
\ead{roxanast75@gmail.com}
\author{Shivakeshavan ~R.G\fnref{fn2}}
\ead{shivak7@ufl.edu}
\author{Pramod P. ~Khargonekar\fnref{fn3}}
\ead{ppk@ece.ufl.edu}
\author{Sachin S. ~Talathi\corref{cor1}\fnref{fn1,fn2,fn4}}
\ead{talathi@ufl.edu}

\cortext[cor1]{Corresponding author}
\fntext[fn1]{Department of Pediatrics, Division of Neurology, University of Florida, Gainesville, FL, USA}
\fntext[fn2]{Department of Biomedical Engineering, University of Florida, Gainesville, FL, USA}
\fntext[fn3]{Department of Electrical Engineering, University of Florida, Gainesville, FL, USA}
\fntext[fn4]{Department of Neuroscience, University of Florida, Gainesville, FL, USA}

\begin{abstract}
Channelrhodopsins-2 (ChR2) are a class of light sensitive proteins that offer the ability to use light stimulation to regulate neural activity with millisecond precision. In order to address the limitations in the efficacy of the wild-type ChR2 (ChRwt) to achieve this objective, new variants of ChR2 that exhibit fast mono-exponential photocurrent decay  characteristics have been recently developed and validated. In this paper, we investigate whether the framework of transition rate model with 4 states, primarily developed to mimic the bi-exponential photocurrent decay kinetics of ChRwt, as opposed to the low complexity 3 state model, is warranted to mimic the mono-exponential photocurrent decay kinetics of the newly developed fast ChR2 variants: ChETA (Gunaydin et al., Nature Neurosci, 13:387-392, 2010) and ChRET/TC (Berndt et al., PNAS, 108:7595-7600, 2011).  We begin by estimating the parameters for the 3-state and 4-state models from experimental data on the photocurrent kinetics of ChRwt, ChETA and ChRET/TC. We then incorporate these models into a fast-spiking interneuron model (Wang and Buzsaki., J Neurosci, 16:6402-6413,1996) and a hippocampal pyramidal cell model (Golomb et al., J Neurophysiol, 96:1912-1926, 2006) and investigate the extent to which the experimentally observed neural response to various optostimulation protocols can be captured by these models. We demonstrate that for all ChR2 variants investigated, the 4 state model implementation is better able to capture neural response consistent with experiments across wide range of optostimulation protocol. We conclude by analytically investigating the conditions under which the characteristic specific to the 3-state model, namely the mono-exponential photocurrent decay of the newly developed variants of ChR2, can occurs in the framework of the 4-state model.
\end{abstract}
\end{frontmatter}
\clearpage

\section{Introduction}
Optogenetics,  an emerging optical neurostimulation technique,  employs light activated ion channels to excite or suppress impulse activity in neurons with high temporal and spatial resolution \citep{Deisseroth:2006fk}. The use of light activation for modulating properties of neurons, an idea that goes back to Nobelist Francis Crick \citep{Crick:1979zr}, has generated great excitement in the biological science community \citep{Kuehn_2010}. Indeed, in recent years there has been a great surge in research efforts using this technology to address fundamental questions in the field of neuroscience \citep{Huber2008,Cardin2009,Bass2010} and neurological disease \citep{Gradinaru:2009kx, Tonnesen:2009vn, Kravitz2010}.

The basic idea behind optogenetic technology is to enable the induction of light activated ion channel proteins (opsins) into neuronal membranes of living animals using techniques from molecular and genetic engineering. The activity of the opsin containing neurons can then be controlled directly via light stimulation. Specifically, experimental strategies have been developed for delivering channelrhodopsin-2 (ChR2), a subfamily of blue light gated-cation opsins, and halorhodopsin (Halo/NpHR), a subfamily of yellow light gated-anion opsins, such that optostimulation can either activate (generate action potential) or deactivate (hyperpolarize) the neuron \citep{Boyden:2005uq,Deisseroth:2006fk,Zhang2007}.   A key merit of this technique is that it offers the ability to regulate neuronal activity with millisecond precision, which in turn allows for fine control of  neuronal activity patterns in the brain region of interest. In addition, these proteins can be engineered to be expressed only in certain types of neurons \citep{Cardin2009}. Thus, this technique offers capability to control neuronal activity with high degree of temporal accuracy in a cell specific manner, a significant advantage over traditional techniques such as electrical stimulation and pharmacological approaches.

Recent {\it in vitro} experiments to characterize the response of neurons expressing the wild-type ChR2 (ChRwt) to various optostimulation protocols have highlighted several limitations in the precision of optogenetic control that can be achieved via optostimulation \citep{Gunaydin2010}. Specifically, short light pulse (2 ms) stimulation of neurons expressing ChRwt can result in the generation of extra spikes; fast-spiking interneurons driven with periodic short light pulse stimulation in the gamma range (30-80 Hz) fail to respond with spiking in the gamma range while higher frequency optostimulation ($>$80 Hz) triggers plateau potential neural response. In order to overcome these limitations, ChR2 mutations with fast photocurrent decay charcteristics have been engineered and validated, which offer enhanced and precise control of neural activity \citep{Gunaydin2010, Berndt2011}.

From a modeling perspective, two transition rate models, a 3-state \citep{Nikolic2006} and a 4-state model \citep{Nikolic2009}, are currently available to mimic the ChRwt photocurrent kinetics. The 4 state model was proposed to specifically capture the bi-exponential decay profile of the ChRwt photocurrent following the termination of the light stimulation protocol \citep{Nikolic2009}.  However, to our knowledge, the bi-exponential photocurrent decay characteristic has not been reported in other studies using ChR2wt or any other ChR2 variant for optogenetic manipulation of neural activity. Therefore, {\it the benefit of enhanced repertoire of dynamical behavior offered by an additional state variable in the transition rate model relative to the increase in the complexity of the model is called into question}. Moreover, it is unclear whether either of these transition rate models can capture the photocurrent kinetics of the recently engineered ChR2 variants \citep{Gunaydin2010,Berndt2011}. Furthermore, no study has yet investigated whether neural activity elicited in model neurons using photocurrents generated by these transition rate models is in agreement with experimental results obtained when various optostimulation protocols are delivered. Given that modern control engineering analysis and design tools depend crucially on mathematical models of the actuation system (ChR2 photocurrents) and the phenomena to be controlled (neural network activity), it is of vital importance to characterize the efficacy of mathematical models of ChR2 photocurrents to reproduce the main features of neural response to various optostimulation protocols.

This study is focused on a systematic investigation of the ability for a 3-state and a 4-state transition rate model respectively, to mimic photocurrent characteristics of ChRwt that exhibits mono-exponential photocurrent decay and the two newly engineered ChR2 variants, namely, ChETA \citep{Gunaydin2010} and ChRET/TC \citep{Berndt2011} in order to determine which of the models can effectively capture the  the dynamical response of the neurons expressing these variants to various optostimulation protocols.  We begin by first identifying the parameters for the 3-state and the 4-state model of each of the three ChR2 variants. We then incorporate these models into two well established neuron models, namely the Golomb model for hippocampal pyramidal cells \citep{Golomb2006} and the Wang-Buszaki model for fast spiking hippocampal interneurons \citep{Buzsaki1996} and study the dynamical response of these neuron models to different optostimulation protocols. Finally, we present analytical results on the conditions under which a characteristic specific to the 3-state model, namely the mono-exponential decay of the ChR2 photocurrent kinetics, can occur in the 4-state model.

The paper is organized as follows: We first summarize the key experimental results incorporated in our development of the 3-state and 4-state transition rate models of ChR2 photocurrent kinetics. We then present the mathematical framework of the transition rate models, which is then followed by the presentation of the neuron models studied in this manuscript. Results from our analysis of the two transition rate models are then presented beginning with results on our estimates of parameter values for the two transition rate models. The discussion section then follows, where we explain the relevance of this study for the development of neural control strategies using optostimulation based actuation systems.

\section{Methods}

\subsection{Experimental data}

The 3- and 4-state transition rate models for each of the ChR2 variant investigated in this paper are designed to match the experimental data currently available in the literature \citep{Gunaydin2010,Berndt2011}. The available data correspond to photocurrent time profiles obtained from whole cell patch clamp recordings from neurons expressing ChR2 in response to prolonged (1 s) optostimulation.

 The following empirically estimated parameters  derived from the time profile of the recorded ChR2 photocurrents  were employed in the development of the two transition rate models: the time $t_{p}$, of peak ChR2 photocurrent ($I_{\text{peak}}$) after the optostimulation pulse was turned on, the time constant $\tau_{\text{in}}$, of the exponential decay of the ChR2 photocurrent from peak to steady state plateau ($I_{\text{plat}}$), the time constant $\tau_{\text{off}}$, of the exponential decay of the photocurrent from plateau to zero when light pulse is turned off, the ratio $R = I_{\text{plat}}/I_{\text{peak}}$ and the value of $I_{\text{peak}}$. The first parameter $t_{p}$, allows us to estimate the time constant $\tau_{\text{rise}}$, of the photocurrent rise (see Appendix, Section 6.4.1 for more details). In addition, the ChR2 photocurrent recovery curve (corresponding to the recovery of the peak current following stimulation with a second light pulse) provides the recovery time constant $\tau_{r}$. The values for these parameters derived from experimental data are summarized in  Table \ref{table1}.

\begin{table*}\small
\begin{center}
    \begin{tabular}{|c|c|c|c|c|c|c|c|c|c|c|}
    \hline
     Experimental Data                 & ChR2 Variant & $t_{p}$ & $\tau_{\text{rise}}(**)$ & $\tau_{\text{in}}$ & $\tau_{\text{off}}$ & $\tau_{r}$ & $R$ & $I_\text{{peak}}$& V$_{\text{hold}}$& $I_{light}$\\
     &&(ms)&(ms)&(ms)&(ms)&(ms)&&(nA)& (mV)&($\frac{\text{mW}}{\text{mm}^{2}}$)\\ \hline
    Gunaydin et al. & ChRwt        & 2.4 & 0.2 & 55.5 & 9.8 & 10700 & 0.4 & 0.848&-100&50 \\                            \citep{Gunaydin2010}           & ChETA       & 0.9 & 0.08 & 15 & 5.2 & 1000 & 0.6 & 0.645&& \\ \hline
    Berndt et al.     & ChRwt        & 2.65 & 0.23 & 9.6(*) & 11.1 & 10700 & 0.27(*) & 0.967&-75&42  \\
                       \citep{Berndt2011}                & ChRET/TC     & 2.17 & 0.18 & 11(*) & 8.1 & 2600 & 0.31(*) & 1.420&& \\
    \hline
    \end{tabular}\vspace{0.2cm}
    \caption{\textbf{Experimental Data.} The values of various time constants and photocurrent quantifiers determined experimentally for ChRwt and the two fast variants ChETA and ChRET/TC are presented. These data have been employed in the evaluation process of the parameters for the 3 and 4 state ChR2 model for each variant. (*) These data have been obtained by direct correspondence with the authors of the experimental study. (**) These data have been analytically evaluated (see Appendix, Section 6.4.1)} \label{table1}
\end{center}
\end{table*}

\subsection{Transition rate models}
ChR2 photocurrent kinetics can be described using the general mathematical framework of transition rate models \citep{Koch_1998}. Such models are described in terms of states $S_{1}\cdots S_{n}$, $n\ge 2$, such that the transitions between any two states is given as:
\bea
\label{eq1s}
S_{i}\overset{r_{ij}}{\underset{r_{ji}}\rightleftharpoons}S_{j}
\eea
where the rates $r_{ij}$ and $r_{ji}$ govern the transition between states $S_{i}$ and $S_{j}$. The fraction of channels in the state $S_{i}$ is denoted by $s_{i}$ and obeys  the following first order differential equation
\bea
\label{eq2s}
\frac{ds_{i}}{dt}&=&\sum_{j=1}^{n}r_{ji}s_{j}-\sum_{j=1}^{n}r_{ij}s_{i}
\eea
such that $\sum_{i=1}^{n}{s_{i}}=1$.

\begin{figure}[htbp]
  \centering
    \includegraphics[scale=0.5]{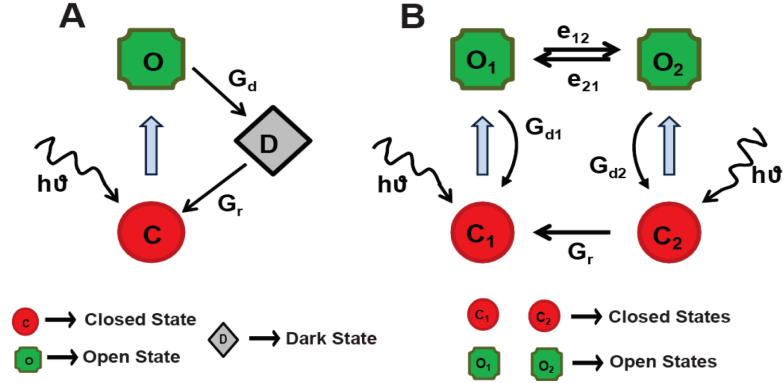}
  \caption{\textbf{Schematic diagram of the 3 and 4-state models.} \textbf{A.} Representation of the 3-state model with an open state $O$, a closed $C$ and a dark state $D$. \textbf{B.} The 4-state model is represented by transitions among two open states ($O_{1}$ and $O_{2}$) and two closed states ($C_{1}$ and $C_{2}$).}
  \label{Fig1}
\end{figure}

In Figure \ref{Fig1}, we present a schematic diagram of the 3-state and the 4-state transition rate models for ChR2 photocurrent kinetics based on the framework described above \citep{Nagel2003,Nikolic2006}.

\subsubsection{3-state transition rate model for ChR2 photocurrent kinetics}

The 3-state model \citep{Nagel2003,Nikolic2006} is schematically described in Figure \ref{Fig1}A. The model simulates the photocurrent kinetics of ChR2 through a set of transitions between three distinct states: closed  $C$, open $O$ and a desensitized state $D$. In the absence of optostimulation, ChR2 molecules are assumed to be in state $C$. Upon illumination with light of appropriate wavelength ($\approx 475$ nm), the ChR2 molecules undergo conformational changes and transition to state $O$, which then spontaneously decays into a closed but desensitized state $D$. ChR2 molecules in state $D$ are not available to photoswitch on optostimulation. Finally, following a prolonged recovery time period, which is  much slower than the time scales involved in the light induced $C\rightarrow O$ and the spontaneous $O\rightarrow D$ transitions, the protein returns to the conformation of the closed state $C$.

The light mediated transition process from the closed state $C$ to the open state $O$ is very fast ($< 1$ ms). The transition from the open state $O$ to the desensitized state $D$ depends on the pH of the media (extracellular solution) and lasts $\approx$10-400 ms \citep{Nikolic2006}. In our modeling framework we do not consider the pH  dependence of the transitions; instead, we will rely on the data reported in Table 1, which has been collected for a media of pH $= 7.5$. If $c$, $o$ and $d$ denote the fraction of ChR2 molecules in each of the three states at any given instant in time, then following from equation \ref{eq2s}, the transition rate model for ChR2 kinetics can be described via the following set of ordinary differential equations:
\begin{eqnarray}\label{eq3s}
\dot{o} &=& P(1-o-d) - G_{d}o\\
\dot{d} &=& G_{d}o - G_{r}d \nonumber
\end{eqnarray}
where, $o+d+c=1$.
The parameters: $P$, the light dependent excitation rate for the $C\rightarrow O$ transition, $G_{d}$, the rate constant for the $O\rightarrow D$ transition and $G_{r}$, the rate constant for the $D\rightarrow C$ transition, are the model parameters to be determined from experimental data summarized in Section 2.1. The ChR2 photocurrent entering the neuron membrane is given as:
\begin{equation} \label{eq4s}
I_{\text{ChR2}} = g_{1}Vo
\end{equation}
where $V$ is the membrane potential  of the neuron expressing ChR2 and $g_{1}$ is the maximal conductance of the ChR2 ion channel.

Closed form analytical expression for $o$ and therefore $I{_\text{ChR2}}$ (under voltage clamp conditions, following from equation \ref{eq1}) under both constant light intensity optostimulation and no light conditions can be obtained (see \citep{Nikolic2006,Nikolic2009} for a complete derivation) from equation \ref{eq3s} as:
\begin{equation}\label{eq5s}
o^{\text{on}}_{\text{ChR2}}(t) = C_{1}e^{-\lambda_{1}t} + C_{2}e^{-\lambda_{2}t} + o_{plat}
\end{equation}

\begin{equation}\label{eq6s}
o^{\text{off}}_{\text{ChR2}}(t) = Ce^{-G_{d}t};
\end{equation}
where the coefficients $C_{1}, C_{2}$, the steady state fraction of ChR2 molecules in open state, $o_{plat}$ and the time decays rate constants $\lambda_{1},\lambda_{2}$($>\lambda_{1}$) depend on the model parameters \{$P,G_{d},G_{r}$\} and the initial conditions \{$o_{0},d_{0}$\} (see Section 6.1 of the Appendix for the formulas). The transition rate parameters \{$G_{d},G_{r}$\} and the photocurrent decay rate constants \{$\lambda_{1},\lambda_{2}$\} are directly obtained from the experimental data on ChR2 photocurrents as:

\begin{equation}\label{eq7s}
G_{d} = \frac{1}{\tau_{off}}; \text{ } G_{r} = \frac{1}{\tau_{r}}; \text{ } \lambda_{1} = \frac{1}{\tau_{\text{in}}}\text{ } \lambda_{2}=\frac{1}{\tau_{\text{rise}}}
\end{equation}
In order to estimate $P$, we use the analytical solution to equation \ref{eq3s} under the light on conditions  to obtain the closed form equation (see Appendix, Sections 6.1 and 6.2):
\begin{equation}\label{eq8s}
P = \lambda_{1} + \frac{G_{r}G_{d}}{\lambda_{1} - G_{r} - G_{d}}
\end{equation}
In Table \ref{table2}, we summarize the values obtained for  the 3-state model parameters for all ChR2 variants from equation \ref{eq7s}, equation \ref{eq8s} and the corresponding experimental data.

\begin{table*}
\begin{center}
    \begin{tabular}{|c|c|c|c|c|c|c|}
    \hline
    Experimental Data           & ChR2 Variant & P (ms$^{-1}$)  & $G_{d}(ms^{-1})$ & $G_{r}(ms^{-1})$ & $g_{1}(*)(IIC)$ & $g_{1}(*)(SIC)$\\ \hline
    Gunaydin \citep{Gunaydin2010} & ChRwt        & 0.0179 & 0.1020 & 9.3458e-005 & 0.07 & 3.687\\ \cline{2-7}
                                       & ChETA       & 0.0651 & 0.1923 & 1e-003 & 0.03314 & 0.7588\\ \hline
    Berndt \citep{Berndt2011}     & ChRwt        & 0.1048 & 0.0901 & 9.3458e-005 & 0.03256 & 3.3728 \\ \cline{2-7}
                                       & ChRET/TC     & 0.0895 & 0.1235 & 3.8462e-004 & 0.06097 & 1.899 \\
    \hline
    \end{tabular}\vspace{0.2cm}
    \caption{\textbf{Parameters of the 3-state model.} The parameters employed in the 3-state model have been determined for ChRwt and the two fast variants ChETA and ChRET/TC using experimental data presented in Table \ref{table1} and the formulas \ref{eq3}-\ref{eq6} from this section. (*)This parameter is obtained by solving the optimization problem of the photocurrent peak by minimizing the error function provided by equation \ref{eq101} for different initial conditions (IIC = Ideal Initial Conditions; SIC = Special Initial Conditions).} \label{table2}
\end{center}
\end{table*}

\subsection{4-state transition rate model for ChR2 photocurrent kinetics}
The 4-state transition rate model is schematically described in Figure \ref{Fig1}B \citep{Nikolic2009}. This model was primarily developed to account for the key experimental findings related to the bi-exponential decay rates for the wild-type ChR2 photocurrent following the termination of prolonged optostimulation pulse, as observed in the experiments of \citep{Ishizuka:2006ly,Nikolic2009}. The 4-state model  simulates the photocurrent kinetics for the light activated ChR2 channel through two sets of intratransitional states: C1$\leftrightarrow$  O1, which is more dark adapted and C2 $\leftrightarrow$ O2, which is more light adapted. In absence of optostimulation, the ChR2 molecules are assumed to be in the closed state C1. In the light adapted state, however, ChR2 molecules are distributed across all four states, with increasing preference to be in state O2 as the duration of optical illumination increases. For a given level of optical excitation, an equilibrium is established between the states O1 and C1. As the duration of illumination increases, the transition to state O2 becomes significant. The O2 state is also in equilibrium with the state C2, which in turn slowly transition back to the state C1 following the termination of the optical signal with time scale of the recovery period.

If the fraction of ChR2 molecules in each of the four states C1, C2, O1 and O2 are given by $c_1$, $c_2$, $o_1$ and $o_2$ respectively, then the four state model for the ChR2 photocurrent kinetics can be represented via the following set of ordinary differential equations:
\begin{eqnarray}\label{eq9s}
\dot{o_{1}} &=& P_{1}s(1-c_{2}-o_{1}-o_{2}) - (G_{d1}+e_{12})o_{1} + e_{21}o_{2}\nonumber\\
\dot{o_{2}} &=& P_{2}sc_{2} + e_{12}o_{1} - (G_{d2}+e_{21})o_{2}\nonumber\\
\dot{c_{2}} &=& G_{d2}o_{2} - (P_{2}s+G_{r})c_{2}\nonumber\\
\dot{s} &=& (S_{0}(\theta) - s)/ \tau_{ChR2}
\end{eqnarray}
such that $c_{1}+c_{2}+o_{1}+o_{2}=1$. The parameters $P_{1}$ and $P_{2}$ are the maximum excitation rates of the first and second closed states, $G_{d1}$ and $G_{d2}$ are the rate constants for the $O1\rightarrow C1$ and the $O2\rightarrow C2$ transitions respectively, $e_{12}$ and $e_{21}$ are the rate constants for $O1\rightarrow O2$ and $O2\rightarrow O1$ transitions respectively and $G_{r}$ is the recovery rate of the first closed state after the light pulse is turned off. We note that, while the photon absorption and the isomerization of the retinal compound in the rhodopsin is almost instantaneous ($\sim 200$ fs \cite{Wang1994,Scho1991}), the conformational change leading to the open state configuration is a slower process. The function $s$ is designed to capture the temporal kinetics of this conformational change in the protein. This function can be provided in an explicit form  \citep{Nikolic2009} or in an equivalent ordinary differential equation form as in \citep{Talathi2011}. Here we adopt the later mathematical description as it provides a clear advantage in implementing various optostimulations protocols. In equation \ref{eq9s}, $S_{0}(\theta) = 0.5(1+\tanh(120(\theta - 0.1)))$ is a sigmoidal function and $\theta(t) = \sum_{i} \Theta(t-t_{i_{on}})\Theta(t-t_{i_{off}})$ describes the optostimulation protocol; $\Theta(x) = 1$ if $x>0$ else $\Theta(x) = 0$ is the Heaviside function while $t_{i_{on}}$ and $t_{i_{off}}$ are the onset, respectively the offset times of the $i_{th}$ optostimulation pulse. The constant $\tau_{ChR2}$ is the activation time of the ChR2 ion channel with typical values on the order of a few milliseconds.

The ChR2 photocurrent is given by:
\begin{eqnarray}\label{eq10s}
I_{\text{ChR2}} &=& V(g_{1}o_{1} + g_{2}o_{2})\nonumber\\
         &=& Vg_{1}(o_{1} + \gamma o_{2});\text{ where } \gamma = \frac{g_{2}}{g_{1}}
\end{eqnarray}

The only parameter of the 4-state model described above that can be identified directly from the experimental data is $G_{r} = 1/\tau_{r}$. Since the complexity of the 4-state model as represented in equation \ref{eq9s} does not allow for estimation of closed form analytical solutions for the 4-state model parameters, $\alpha=\{P_{1},P_{2},G_{d1},G_{d2},e_{12},e_{21},\gamma, g_{1}, \tau_{ChR2}\}$, we employ the following empirical estimation procedure to estimate these parameters.

First, we establish an empirical formula to describe the photocurrent profile obtained from the  experimental data for the ChR2 photocurrent measurements obtained for a long term (1s) continuous optostimulation ($I_{\text{emp1}}$) pulse and for a short term (2ms) optostimulation ($I_{\text{emp2}}$) pulse as follows:
\begin{eqnarray}\label{eq11s}
I_{\text{emp1}}(t) &=& I_{\text{peak}}\left(1-e^{-\frac{(t-t_{\text{on}})}{\tau_{\text{rise}}}}\right) \nonumber \\
&&\Theta(t-t_{\text{on}})\Theta((t_{on}+t_{p})-t) \nonumber \\
&+& I_{\text{peak}}\left(R + (1-R)e^{-\frac{(t-(t_{on}+t_{p}))}{\tau_{\text{in}}}}\right) \nonumber \\
&&\Theta(t-(t_{on}+t_{p}))\Theta(t_{\text{off}}-t) \nonumber \\
  &+& RI_{\text{peak}}e^{-\frac{(t-t_{\text{off}})}{\tau_{\text{off}}}}\Theta(t-t_{\text{off}})
\end{eqnarray}
\begin{eqnarray}\label{eq12s}
I_{\text{emp2}}(t) = I_{\text{peak}} \Bigg[ (1-e^{-\frac{(t-t_{\text{on}})}{\tau_{\text{rise}}}})\\ \nonumber %
\Theta(t-t_{\text{on}})\Theta(t_{p}-t) + e^{-\frac{(t-(t_{on}+t_{p}))}{\tau_{\text{off}}}}\Theta(t-t_{\text{p}}) \Bigg]
\end{eqnarray}
where $\Theta$ is the Heaviside step function, $t_{\text{on}}$ is the time when the optostimulation pulse is turned on, $t_{p}$ is the time when ChR2 photocurrent reaches its peak value after the light is turned on and $t_{\text{off}}$ is the time when the optostimulation pulse is turned off.

The first term in equation \ref{eq11s} describes the rise of ChR2 photocurrent from zero to peak after the initiation of the optostimulation pulse. The second term simulates the decay of the ChR2 photocurrent from the peak to the steady state $I_{\text{plat}}$ and the last term accounts for the decay of the photocurrent from  $I_{\text{plat}}$ to zero after the optostimulation pulse is turned off. Similarly, in equation \ref{eq12s}, the first term accounts for the rise of the photocurrent from zero and approaching peak value following brief optostimulation and the second term describes the exponential decay to zero from the maximum photocurrent value reached during the light on phase of optostimulation. We note that in writing equations \ref{eq11s} and \ref{eq12s}, we have assumed that $2<\tau_{\text{rise}}<<1000$ ms, a valid assumption for all the ChR2 variants that we have studied here.

We begin with an initial set of parameter values \{$\alpha$\} (that obey the constraints detailed below) and numerically solve equation \ref{eq9s} to obtain $I_{\text{ChR2}}(t,\alpha)$ using equation \ref{eq10s} with $V=V_{\text{hold}}$ for both long term and short term optostimulation protocols. We then evaluate the following objective function:
\begin{equation}\label{eq13s}
C(\alpha)= E_{1}(\alpha) + E_{2}(\alpha) +E_{3}(\alpha)
\end{equation}
where:
\begin{eqnarray}\label{eq14s}
E_{1}(\alpha) = 100*\sqrt{\frac{1}{T_{1}}\int_{0}^{T_{1}}\left[I_{\text{ChR2}}(t,\alpha) - I_{\text{emp1}}(t)\right]^2 dt}\nonumber \\
E_{2}(\alpha) = 100*\sqrt{\frac{1}{T_{2}}\int_{0}^{T_{2}}\left[I_{\text{ChR2}}(t,\alpha) - I_{\text{emp2}}(t)\right]^2 dt}\nonumber\\
\end{eqnarray}
are the Root Means Square Deviations (RMSD) \% errors of the photocurrent elicited by $T_{1}=1$ s and $T_{2}=2$ ms optostimulation pulse relative to their empirical profile and
\begin{equation}\label{eq15s}
E_{3}(\alpha) = 100*\frac{|I_{\text{peak}_{\text{ChR2}}}(\alpha) - I_{\text{peak}_{\text{emp1}}}|}{I_{\text{peak}_{\text{emp1}}}}
\end{equation}
is the percent error of the photocurrent peak achieved with prolonged optostimulation.

We solve the constrained optimization problem: $\underset{\alpha}{\operatorname{argmin}}\text{ } C(\alpha)$, using a global search method (genetic algorithm) under the following constraints: 1) all parameters must be positive numbers, 2) for given light intensity (see table \ref{table1}), the values for the parameters $P_{1}$ and $P_{2}$ must not exceed a maximum value $P_{max}$ ($P_{1}<P_{max}$ and $P_{2}<P_{max}$) (evaluated using formula (38) from Appendix for maximum quantum efficiency, $\epsilon = 1$, and minimum degree of light absorbtion and scattering, $w_{loss}=1$). 

The parameters are further optimized using a constrained local gradient descent algorithm (fmincon in Matlab). In Table \ref{table3}, we present the values for \{$\alpha$\} obtained using the procedure described above for each of the ChR2 variants.

\begin{table*} \footnotesize 
\begin{center}
    \begin{tabular}{|c|c|c|c|c|c|c|c|c|c|c|c|}
    \hline
    Experimental           & ChR2 & $P_{1}$ & $P_{2}$ & $G_{d1}$ & $G_{d2}$ & $e_{12}$ & $e_{21}$ & $G_{r}$ & $\tau_{\text{ChR2}}$ & $\gamma$ & $g_{1} (*)$\\
    Data&Variant&(ms$^{-1}$)&(ms$^{-1}$)&(ms$^{-1}$)&(ms$^{-1}$)&(ms$^{-1}$)&(ms$^{-1}$)&(ms$^{-1}$)& (ms$^{-1}$)&\\ \hline
    Gunaydin & ChRwt          & 0.0641 & 0.06102 & 0.4558 & 0.0704 & 0.2044 & 0.0090 & 9.3458e-005 & 6.3152 & 0.0305 & 0.1136 \\
    \citep{Gunaydin2010} & ChETA       & 0.0661 & 0.0641 & 0.0102 & 0.1510 & 10.5128 & 0.0050 & 1e-003 & 1.5855 & 0.0141 & 0.8759 \\ \hline
    Berndt & ChRwt            & 0.1243 & 0.0125 & 0.0105 & 0.1181 & 4.3765 & 1.6046 & 9.3458e-005 & 0.504 & 0.0157 & 0.098 \\
    \citep{Berndt2011}   & ChET/TC     & 0.1252 & 0.0176 & 0.0104 & 0.1271 & 16.1087 & 1.0900 & 3.8462e-004 & 0.3615 & 0.0179 & 0.5599 \\
    \hline
    \end{tabular}\vspace{0.2cm}
    \caption{\textbf{Parameters of the 4-state model.} The parameters of the 4-state model have been determined for ChRwt and the two fast variants ChETA and ChRET/TC using experimental data (see Table \ref{table1}) and search algorithms to best fit the profile photocurrent provided by equations \ref{eq11s} and \ref{eq12s}. (*) This parameter is considered a variable in Section 3.2.} \label{table3}
\end{center}
\end{table*}

\subsection{Neuron Models}
The neural response to optostimulation is evaluated using two neuron models: 1) A single compartment fast spiking interneuron model of Wang and Buzsaki (WB) \citep{Buzsaki1996} and 2) A single compartment hippocampal pyramidal neuron model of Golomb et al., (Gol) \citep{Golomb2006}. Each neuron model with the addition of light dependent ChR2 ion channel current  $I_{\text{ChR2}}$ can be described using a system of differential equations of the form:
\bea\label{eq16s}
c\dot{V}&=&I_{\text{DC}}-I_{g}(V,{\bf n})-I_{\text{ChR2}}(V,{\bf s}) \nonumber \\
\dot{{\bf n}}&=&{\bf G}(V,{\bf n})\nonumber \\
\dot{{\bf s}}&=&{\bf H}(V,{\bf s})\nonumber \\
{\bf m_{\infty}}&=&{\bf U}(V)
\eea
where $V$ is the membrane voltage, ${\bf n}\in[0,1]$ and ${\bf m_{\infty}}\in[0,1]$  are the vectors of gating variables for the ion channels present on the neural membrane with finite rise-time constant and instantaneous rise-time kinetics respectively; , ${\bf s}\in[0,1]$ is the vector of transition states of ChR2, $I_{g}$ is the sum of the membrane ion channel currents, $I_{\text{ChR2}}$ is the light activated ChR2 ion channel current, $c$ is the membrane capacitance and $I_{DC}$ is the constant DC bias that controls the excitability of the neuron.\\
For the WB  model the membrane ion currents are, $I_{g}=\{I_{\text{Na}},I_{\text{K}},I_{\text{L}}\}$, where
\begin{eqnarray}\label{eq17s}
I_{Na} &=& g_{Na}m_{\infty}^{3}h(V-E_{Na})\nonumber \\ I_{K} &=& g_{K}n^{4}(V-E_{k})\nonumber \\ I_{L} &= &g_{L}(V-E_{L})
\end{eqnarray}
The gate variables are $\{n,h,m_{\infty}\}$, where
\begin{eqnarray}\label{eq18s}
\frac{dh}{dt} &=& \phi(\alpha_{h}(1-h) - \beta_{h}h)\nonumber \\ \frac{dn}{dt} &=& \phi(\alpha_{n}(1-n) - \beta_{n}n)\nonumber \\ m_{\infty} &=& \frac{\alpha_{m}}{\alpha_{m}+\beta_{m}}
\end{eqnarray}
The functional form for the membrane voltage dependent rate constants \{$\alpha_{m},\beta_{m},\alpha_{h},\beta_{h},\alpha_{n},\beta_{n}$\} are gives as:
\bea\label{eq19s}
\alpha_{m}(V)&=& \frac{-0.1(V+35)}{(\exp(-0.1(V+35))-1)} \nonumber \\
\beta_{m}(V) &=& 4 \exp(-(V+60)/18)  \nonumber \\
\alpha_{h}(V) &=& 0.07(\exp(-(V+58))/20) \nonumber \\
\beta_{h}(V) &=& 1/\exp(-0.1(V+28)+1) \nonumber \\
\alpha_{n}(V) &=& \frac{-0.01(V+34)}{(\exp(-0.1(V+34))-1)}  \nonumber \\
\beta_{n}(V) &=& 0.125 \exp(-(V+44)/80)
\eea
The model parameters are: $E_{Na} = 55$ mV, $E_{K} = -90$ mV, $E_{L} = -65$ mV, $g_{Na} = 35$ mS/cm$^{2}$, $g_{K} = 9$ mS/cm$^{2}$, $g_{l} = 0.1$ mS/cm$^{2}$, $c = 1$ $\mu$F/cm$^{2}$, $\phi=5$. For all the simulations, we set $I_{DC}=-0.51 \mu$A/cm$^{2}$, which ensures that the resting neuron membrane voltage is $V_{\text{rest}}=-70$ mV in absence of optostimulation.

For the Gol model the membrane ion currents are, $I_{g}=\{I_{\text{Na}},I_{\text{Kdr}},I_{\text{L}},I_{\text{A}},I_{\text{M}}\}$, where
\begin{eqnarray}\label{eq20s}
I_{Na} &=& g_{Na}m_{\infty}^{3}h(V-E_{Na}) \nonumber \\I_{L} &=& g_{L}(V-E_{L})\nonumber \\I_{Kdr} &=& g_{Kdr}n^{4}(V-E_{K}) \nonumber \\
I_{NaP} &=& g_{NaP}p_{\infty}^{3}h(V-E_{Na})\nonumber \\I_{A} &=& g_{A}a_{\infty}^3b(V-E_{K})\nonumber \\ I_{M} &=&g_{M}z(V-E_{K})
\end{eqnarray}
The gate variables are $\{n,h,b,z,m_{\infty},p_{\infty},a_{\infty}\}$, where
\begin{eqnarray}\label{eq21s}
\dot{h} &=& \phi*(\Gamma(V,\theta_{h},\sigma_{h})-h)/(1.0+7.5*\Gamma(V,\tau_{th},-6.0))\nonumber \\ \dot{b} &=& (\Gamma(V,\theta_{b},\sigma_{b})-b)/\tau_{b}\nonumber \\
\dot{n} &=& \phi*(\Gamma(V,\theta_{n},\sigma_{n})-n)/(1.0+5.0*\Gamma(V,\tau_{tn},-15.0))\nonumber \\
\dot{z}& =& (\Gamma(V,\theta_{z},\sigma_{z})-z)/\tau_{z}\nonumber \\
m_{\infty} &=& \Gamma(V,\theta_{m},\sigma_{m})\nonumber \\
  p_{\infty} &=& \Gamma(V,\theta_{p},\sigma_{p})\nonumber \\
    a_{\infty} &=& \Gamma(V,\theta_{a},\sigma_{a})
\end{eqnarray}
with $\Gamma(V,\theta,\sigma) = \frac{1}{1+\exp\left(\frac{-(V-\theta)}{\sigma}\right)}$ and the parameters: $
\theta_{m} = -30; \quad \sigma_{m} = 9.5;\quad  \theta_{p} = -47; \quad \sigma_{p} = 3; \quad \theta_{h} = -45;\quad  \sigma_{h} = -7; \quad
\theta_{n} = -35; \quad \sigma_{n} = 10;\quad  \theta_{a} = -50;\quad  \sigma_{a} = 20; \quad \theta_{b} = -80;\quad  \sigma{b} = -6; \quad
\theta_{z} = -39;\quad  \sigma_{z} = 5;\quad  \tau_{b} = 15;\quad  \tau_{z};\quad \tau_{th} = -40.5; \quad \tau_{tn} = -27.
$

The remaining model parameters are: $ E_{Na} = 55 \text{ mV}; \quad E_{K} = -90 \text{ mV}; \quad E_{L} = -70\text{ mV}; \quad \phi = 10;\quad  g_{Na} = 35 \text{ mS/cm}^{2}; \quad g_{NaP} = 0; \quad g_{Kdr} = 6 \text{ mS/cm}^{2}; \quad g_{L} = 0.05 \text{ mS/cm}^2;\quad g_{A} = 1.4 \text {mS/cm}^{2}; \quad g_{M} = 1 \text{mS/cm}^{2}$. For all the simulations with the Gol model, we set $I_{DC}=0.12 \mu$A/cm$^{2}$, which ensures that the resting neuron membrane voltage is $V_{\text{rest}}=-70$ mV in absence of optostimulation.

\textbf{Numerical Methods:} All the simulations are performed using the forth order Runge-Kutta method with a time step of 0.05 ms implemented in Matlab R2011b. The matlab codes will be made available for download at ModelDB.

\section{Results}
\subsection{Modeling the response of non-excitable cells to optostimulation}
\begin{figure*}[htbp]
  \centering
    \includegraphics[scale=0.7]{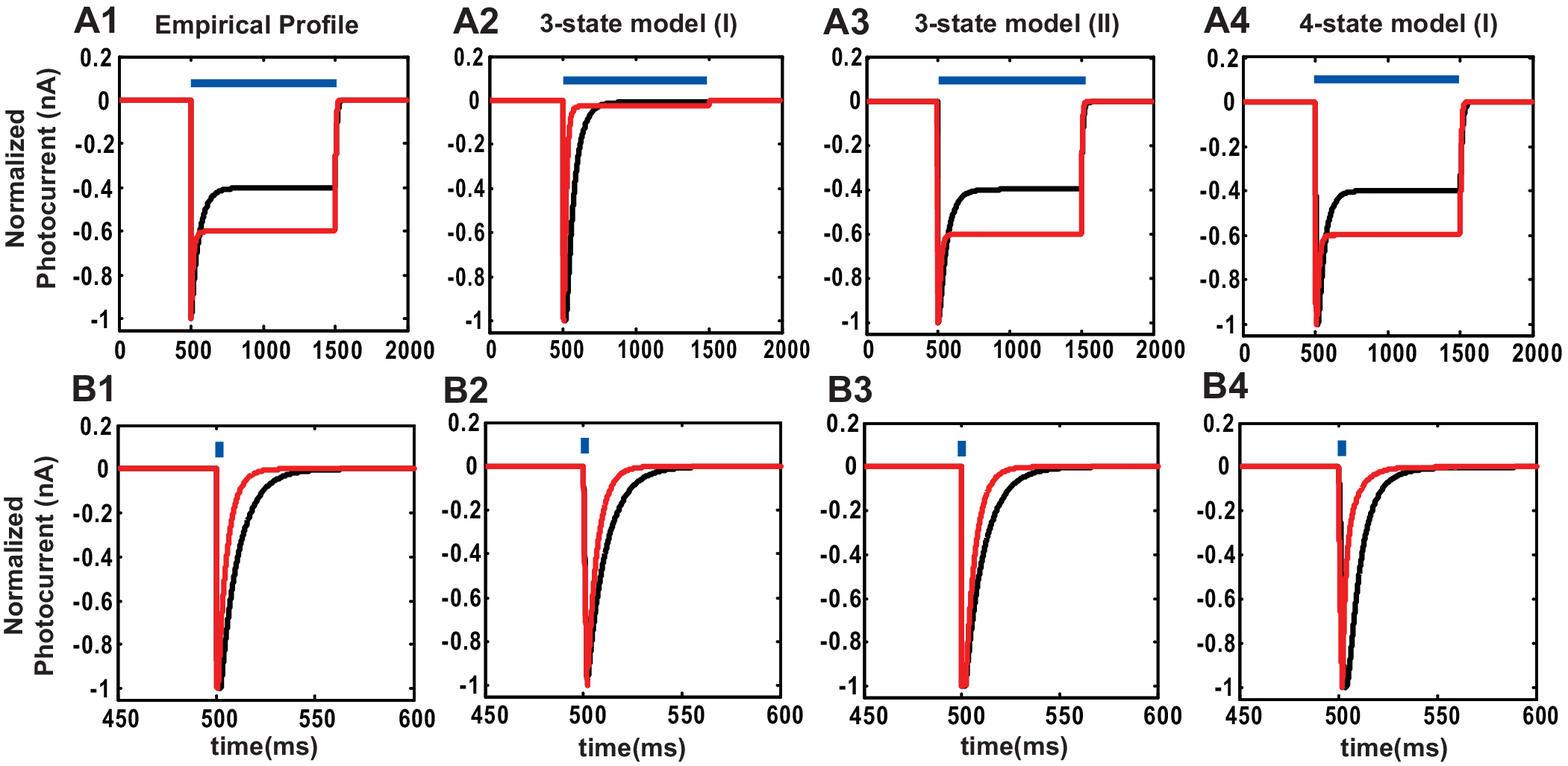}
  \caption{\textbf{Comparison example of the ChR2 photocurrent elicited by 1s and 2 ms optostimulation}. \textbf{A1} and \textbf{B1}. Empirical data profile constructed for ChRwt(black) and ChETA(red) variant using equations \ref{eq9} and \ref{eq91} as well as the experimental data provided in Table \ref{table1} is displayed for a continuous 1s optostimulation (\textbf{A1}) and for a brief 2 ms optostimulation (\textbf{B1}). The two normalized experimental photocurrent profiles are matching the experimental results reported in \citep{Gunaydin2010} Figure 3c and Figure 2d. \textbf{A2} and \textbf{B2}. Photocurrent generated by the 3-state model starting from ideal initial conditions ((I): $C(0) = 1; O(0) = 0; D(0) = 0.$ ) for ChRwt and ChETA variant for 1s (\textbf{A2}) respectively 2ms (\textbf{B2}) optostimulation. \textbf{A3} and \textbf{B3}. Photocurrent generated for ChRwt in comparison with ChETA, by the same 3-state model but starting from special initial conditions ((II): ChRwt: $C(0) = 0.0132; O(0) = 0.0023; D(0) = 0.9845$; ChETA: $C(0) = 0.0251; O(0) = 0.0085; D(0) = 0.9664.$) evaluated in Appendix, Section 6.3. \textbf{A4} and \textbf{B4} Photocurrent generated for the same ChR2 variants using the 4-state model with ideal initial conditions ((I): $C_{1}(0) = 1; O_{1}(0) = 0; O_{2} = 0.; C_{2}(0) = 0.$ ) ). For a similar example comparison of ChR2 photocurrent elicited by ChRwt and ChRET/TC see Figure 10 in Appendix.}
  \label{Fig2}
\end{figure*}

We begin by examining the extent to which the 3-state and the 4-state transition rate models can capture the main features of the ChR2 photocurrent elicited in a non-excitable (oocyte) cell by a prolonged (1s) optostimulation protocol. For the 3-state model, using the parameters obtained in Table \ref{table2} and assuming that all ChR2 molecules are in the dark adapted closed state $C$ prior to the optostimulation, we find that,\emph{ independent of the ChR2 variant}, the 3-state model is unable to reproduce an important feature of experimentally measured ChR2 photocurrent, namely the plateau to peak ChR2 photocurrent ratio $R$ (Figure \ref{Fig2}, panels A2 and Figure 10 panel A2 from Appendix). Similar findings have been reported for ChRwt photocurrent kinetics resulting from a 3-state model \citep{Nikolic2006}. In order to remedy this discrepancy, Nikolic et al., \citep{Nikolic2006} hypothesized that the recovery rate $G_{r}$ for the ChR2 molecules to transition from the desensitized state $D$ to the dark adapted closed state $C$ is a function of light intensity. While this hypothesis is not implausible as stated in \citep{Nikolic2009}, to the best of our knowledge, it has not been yet supported by any experimental evidence.

We next investigate whether the feature $R$, of ChR2 photocurrent,  can be captured using a different set of initial conditions. We note that, the dependence of the 3-state model solution on the initial condition of the state variables was not evaluated in the original formulation of the 3-state model by \citep{Nikolic2006}. In order to address this question, we analytically evaluate the set of initial conditions required to correctly reproduce the ratio $R$ (see Section 6.3 of Appendix). We find that the initial conditions that generate experimentally observed ChR2 photocurrents with reproducible features, correspond to the case when all the ChR2 molecules are in the desensitized state $D$ at the beginning of the optostimulation protocol, suggesting the possibility that the experiments might have been carried out under non-ideal conditions, i.e, all ChR2 molecules have not relaxed back to the closed state before the prolonged optostimulation protocol was run (Figure \ref{Fig2}, panels A3 and Figure 10, panel A3 from Appendix). 

\begin{figure*}[htbp]
  \centering
    \includegraphics[scale=0.5]{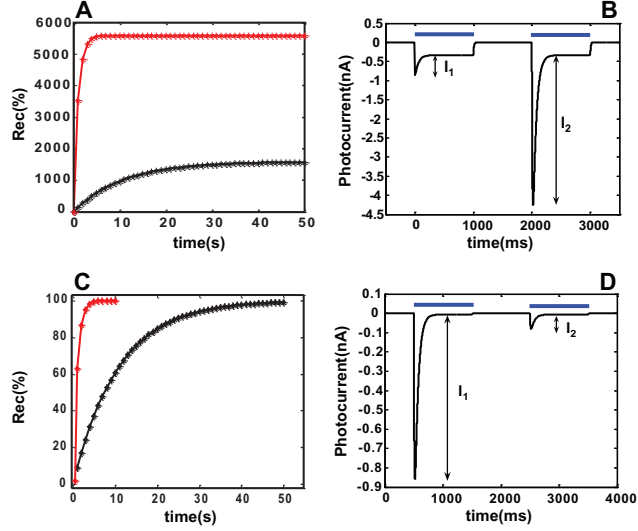}
  \caption{\textbf{Example of recovery curves in the 3 state model}  A. and C. Recovery curves have been evaluated for ChRwt (black) and ChETA (red) for special initial conditions  (A) respectively for ideal initial conditions (C) at the beginning of the optostimulation  protocol. B Example of photocurrent curve obtained for ChRwt when the protein is in special initial condition at the beginning of the stimulation and two pulses of 1 second each are  applied with a dark period of 1 s in between. The recovery ($Rec(\Delta t) = 100*\Delta l_{2}/\Delta l_{1}$) is inconsistent with the experimental observations.  D.  Similar example of photocurrent curve as in B obtained when the protein is in ideal initial conditions at the beginning of stimulation. The recovery is consistent with the experimental observations but the ratio $R$ of photocurrent peak to steady state is not satisfied.}
  \label{Rec_Figure}
  \end{figure*}

In order to determine whether this is indeed the case, we conduct the following simulation experiment: we begin with initial conditions for the 3-state model in the desensitized state $D$. We then present two prolonged optostimulation pulses of 1 s in duration each, separated by time $0 < \Delta t < 50$ s. We then calculate the \% recovery $Rec(\Delta t) = 100* \Delta l_{2}/ \Delta l_{1} $ where $\Delta l_{1,2}$ are the differences between the peak and steady state photocurrents for the two optostimulation pulses respectively. In Figure \ref{Rec_Figure} A and B we present these results in comparison with the corresponding results obtained for simulations started with ideal initial conditions (see Figure \ref{Rec_Figure} C and D). It is clear from Figure \ref{Rec_Figure}B that the photocurrent profile produced by the 3-state model beginning with all ChR2 molecules in the desensitized state $D$ exhibits a recovery percent grater than 100 ($Rec(\Delta t) >100$ for $\Delta t > 0$) . When the 3-state model is simulated with the special initial condition of all ChR2 molecules being in the desensitized state, at the end of first optostimulation protocol, more ChR2 molecules are present in the closed state and ready to maximally respond to an impending optostimulation protocol resulting in  $Rec(\Delta t)>100$. This finding is in stark contrast with the experimental results of Gunaydin et al., \citep{Gunaydin2010}. The experimentally observed recovery curve for ChR2 photocurrents can be reproduced if we assume that all the ChR2 molecules at the beginning of the experiment are in the dark adapted closed state $C$ as shown in Figure \ref{Rec_Figure}C and \ref{Rec_Figure}D. 

Following from these observations, we conclude that while the 3-state model can exhibit ChR2 photocurrent characteristics for a single prolonged optostimulation pulse under special circumstances, it is not a suitable model to mimic the transient photocurrent dynamics in response to multiple optostimulation pulses. We note that the key reason presented in \citep{Nikolic2009} for the unsuitability of 3-state model was low model complexity with insufficient model parameters to capture all the essential features of ChR2 photocurrent profile,  specifically, the bi-exponential photocurrent decay. Our findings on the other hand suggest that for all the ChR2 variants analyzed here, in terms  of model complexity the 3-state model is sufficient to capture all the essential features of ChR2 photocurrent profile. However, this requires non-equilibrium initial distribution of ChR2 molecules across the 3-states in dark adapted conditions. Further support for our claim of the unsuitability of 3-state model will be presented in Section 3.2.

\begin{figure*}[htbp]
  \centering
    \includegraphics[scale=0.5]{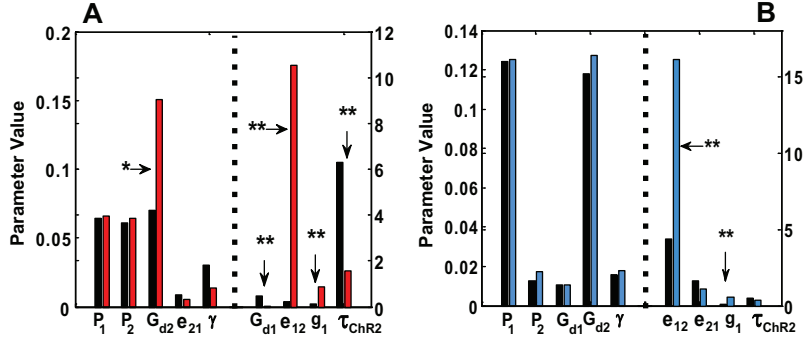}
  \caption{\textbf{Comparison between the coefficients of the 4 state model.} \textbf{A.} Comparison between the 4-state model coefficients evaluated for ChETA (red) and ChRwt (black). \textbf{B.} Similar comparison between the coefficients of ChRwt (black) and the fast variant ChRET/TC (blue). Because of significant differences among the values of the parameters, two scales have been used to visualize their comparison. In both panels, the parameters to the right of the dashed line obeys the scale on the right hand side while the parameters to the left of the dashed line obeys the left hand side scale; difference between the corresponding parameter values for ChRwt and the variant (ChETA or ChRET/TC) larger than $50\%$ but smaller than 3 times the value are indicated with one star and differences larger than 3 times the value are indicated by two stars. Actual values for all parameters are reported in Table \ref{table3}.}\label{FigParam}
\end{figure*}

In the case of the 4 state model, we are able to employ the strategy described in Section 2.3 to identify the set of model parameters that yield ChR2 photocurrents matching the experimental recordings for an ideal set of initial conditions i.e., all the ChR2 molecules are in the dark adapted closed state $C_{1}$ satisfying the initial conditions $c_{1}=1$ and $o_{1}=o_{2}=c_{2}=0$. The following general conclusions can be drawn from the estimated 4-state model parameters (see Figure \ref{FigParam} for a visual comparison of these values): 1) For all the variants $P_{1}>P_{2}$, which underscores the idea that ChR2 molecules exhibit two distinct dark-light adapted closed states with the closed state $C_{1}$ being more light sensitive than the closed state $C_{2}$ to optostimulation as was suggested by Nikolic et al., \citep{Nikolic2009}. 2) The fast transition of ChR2 photocurrents from the peak value to steady state for both variants relative to ChRwt observed in the experiments of Gunaydin et al., \citep{Gunaydin2010} and Berndt et al. \citep{Berndt2011} are explained via significant differences in the transition rate parameters $G_{d1}$ and/or $e_{12}$ across the two variants. Specifically, $G_{d1}^{\text{ChRwt}}>>G_{d1}^{\text{ChETA}}$, $e_{12}^{\text{ChRwt}}<<e_{12}^{\text{ChETA}}$ and $e_{12}^{\text{ChRwt}}<<e_{12}^{\text{ChRET/TC}}$, indicating that the transition from the dark adapted open state $O_{1}$(which accounts for the peak ChR2 photocurrent) into the light adapted open state $O_{2}$ (which accounts for the steady state ChR2 photocurrent) is the preferred mode of operation for the ChETA variant over the situation when the dark adapted open state transitions to the dark adapted closed state. 3) Both variants exhibit higher conductance (quantified here by the parameter $g_{1}$) relative to ChRwt. While in the case of ChETA the expected increase in the maximum amplitude of the photocurrent is suppressed by an enhanced value of the parameter $G_{d2}$ (controlling the decay of the second open state), in the case of the ChRET/TC variant this suppression does not occur (as $G_{d2}^{\text{ChRwt}} \sim G_{d2}^{\text{ChRET/TC}}$) which explains the significantly enhanced photocurrent peak reported for this variant \citep{Berndt2011}. Finally, 4) we notice that both variants show decreased values of the time constant of the channel opening relative to ChRwt ($\tau_{ChR2}^{ChETA} \ll \tau_{ChR2}^{ChRwt}$ and $\tau_{ChR2}^{ChRET/TC} < \tau_{ChR2}^{ChRwt}$) suggesting that the time scale of the conformational changes leading to the channel opening is significantly faster in the newly developed variants relative to ChRwt.

It is important to note that for the ChR2 photocurrent profile generated using either  3-state and 4-state models, the parameter $g_{1}$, expressing the overall conductance of the ChR2 variant, controls the maximum value of the photocurrent amplitude and is independent of other characteristics of the photocurrent dynamics governed by the transition rate models such as the activation ($\tau_{rise}$), inactivation ($\tau_{in}$) and deactivation ($\tau_{off}$) time constants and the ratio $R$ of peak to steady photocurrent amplitude. This parameter therefore, can capture the degree of expression of the ChR2 variant in the membrane of the neural cell and is assumed to be variable across experiments.

\subsection{Modeling the response of excitable cells to optostimulation}
\begin{figure*}[htbp]
  \centering
    \includegraphics[scale=0.7]{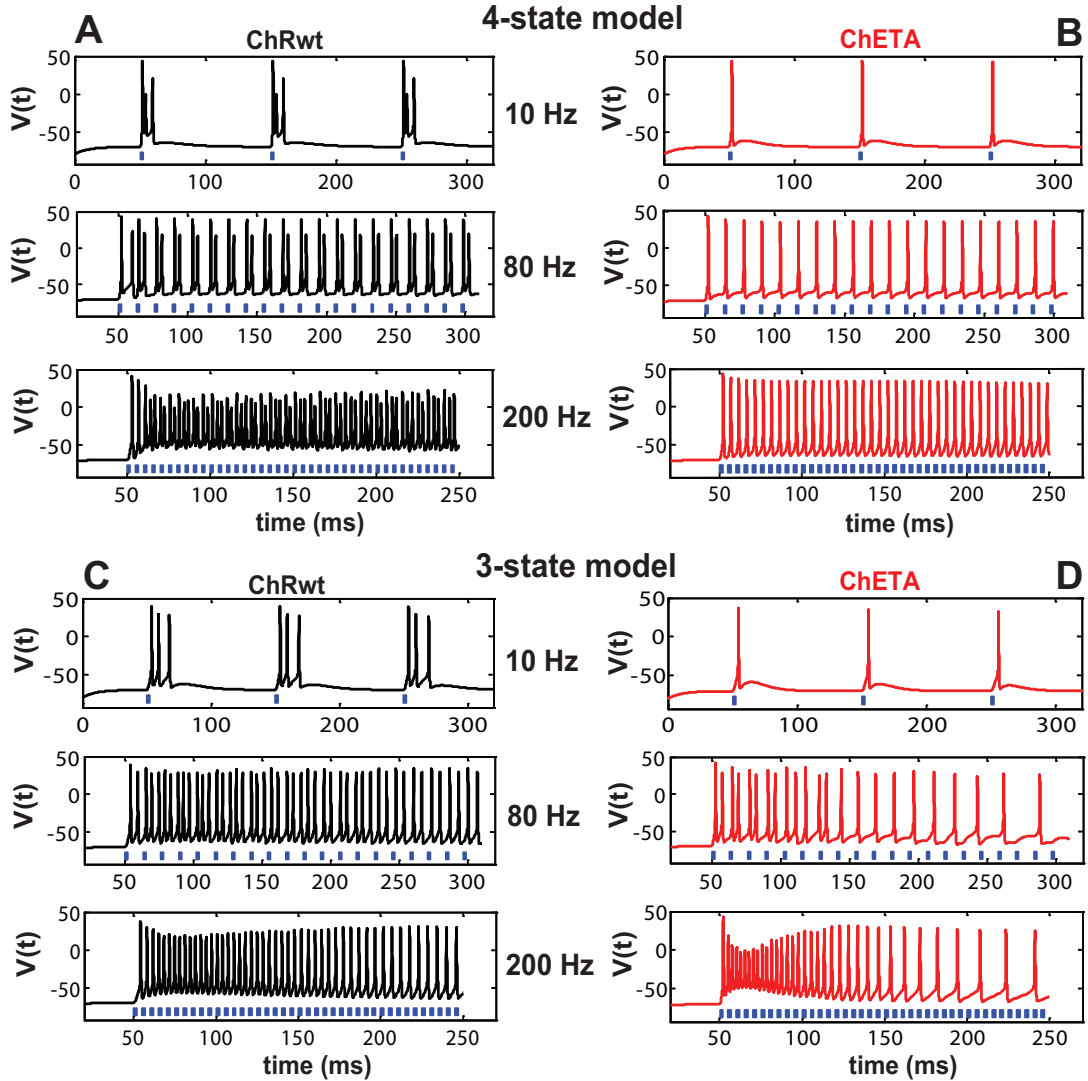}
  \caption{\textbf{Comparison between the simulated interneural response for different optostimulation protocols when 4-state respectively 3-state model accounts for ChR2 dynamics.} \textbf{A.} Simulated neural response to 3 (upper panel), 20 (middle panel) and 40 (lower panel) stimuli of 2 ms each presented at 10, 80 and 200 Hz optostimulation frequency obtained when the 4-state model is employed to account for ChRwt dynamics; the overall conductance was $g_{1} = 100$ (upper panel), $g_{1} = 40$ (middle panel), respectively $g_{1} = 20$ (lower panel). \textbf{B.} The response to the same optostimulation protocol is simulated in interneuron model expressing ChETA; the overall conductance was $g_{1}=70$ for all simulations. These results are in excellent agreement with experimental data presented in \citep{Gunaydin2010} Figure 4 a). \textbf{C.} and \textbf{D.} Results obtained for the same optostimulation protocol when the 3-state model is representing the ChRwt (\textbf{C}) and ChETA (\textbf{D}) kinetics. The overall conductance has the following values: in \textbf{C} $g_{1} = 4.8$ (upper and middle panel), $g_{1} = 4$ (lower panel); in \textbf{D} $g_{1} = 1.2$ (upper panel), $g_{1} = 2.2$ (middle panel) and $g_{1} = 4$ (lower panel). The differences between the results obtained for the 3 and 4- state model are discussed in detail in the text.}
  \label{Fig3}
\end{figure*}

We now investigate whether the two transition rate models can capture the experimental results on the neural response to various optostimulation protocols \citep{Gunaydin2010, Berndt2011}. We begin our investigation by incorporating the 3- and 4-state ChR2 photocurrent transition rate models into the WB model \citep{Buzsaki1996}. In Figure \ref{Fig3}, we compare our simulation results with experimental data on neural activity elicited in the fast spiking hippocampal interneuron models expressing ChRwt and ChETA for various short duration ($<<$1 s) optostimulation protocols \citep{Gunaydin2010}. We make the following observations: for low frequency optostimulation protocols ($<20$ Hz) we find that both the 3- and 4-state models are able generate neural response from the WB model neuron consistent with the experimental recordings. For instance, for 10 Hz optostimulation frequency, the simulated neural response using the 4-state model for ChR2 kinetics (see Fig. \ref{Fig3}A and \ref{Fig3}B the top panels) matches very well the experimental observations for the same optostimulation protocol (Figure 4a in \citep{Gunaydin2010}). Similar results are obtained when the 3-state model is employed to model the ChR2 photocurrent kinetics (see top panels in Figure \ref{Fig3}C and \ref{Fig3}D).

For higher frequency ($>40$ Hz) optostimulation protocols, significant differences in the neural activity elicited by the two transition rate models are revealed. In particular, the 3-state model is not able to capture the bursting features observed in the neural activity of interneurons expressing ChRwt in response to 80 Hz optostimulation protocol (see Figure \ref{Fig3}C middle panel in comparison with Figure \ref{Fig3}A middle panel and Figure 4a middle-left panel in \citep{Gunaydin2010}); neither is the 3-state model able to capture the neural activity of interneurons expressing ChETA. For the later case, the neural response exhibits features inconsistent with the experimental findings including extra spikes at the beginning of the stimulation protocol and missed spikes at the end of the stimulation protocol (see Fig. \ref{Fig3}D middle panel). These inconsistencies are not observed when the 4-state model is implemented in the WB model neuron (Figure \ref{Fig3}B middle panel compared to Figure 4a middle-right panel in \citep{Gunaydin2010}).

For very high frequency ($>150$ Hz)  optostimulation protocols, the 4-state model is again able to  better capture the features of neural response for both the ChRwt and ChETA mutant (Figure \ref{Fig3}A and \ref{Fig3}B bottom panels in comparison with Figure 4a bottom panel from \citep{Gunaydin2010}). For instance, the plateau potential observed for neurons expressing ChRwt is well reproduced by our simulation results produced using the 4-state model (see Figure \ref{Fig3}A bottom panel) and at the same time our simulation experiments match the experimental findings from the neuron expressing ChETA in the sense that the plateau behavior is absent (Figure \ref{Fig3}B bottom panel). On the other hand,  the simulated neural response of the WB model neuron with a 3-state model implementation of the ChRwt and ChETA variant shows several artifacts such as, a decrease of the plateau potential when a larger number of stimulation pulses is presented to the neuron expressing ChRwt, depolarization block at the beginning of the stimulation protocol and inconsistencies in the  interspike intervals observed in neurons expressing ChETA (Figure \ref{Fig3}C and \ref{Fig3}D bottom panels). These inconsistencies are primarily driven by the limitations of the 3-state model to mimic the recovery rate in response to multiple short duration optostimulation light pulses, as explained in the Section 3.1.

\begin{figure*}[htbp]
  \centering
    \includegraphics[scale=0.6]{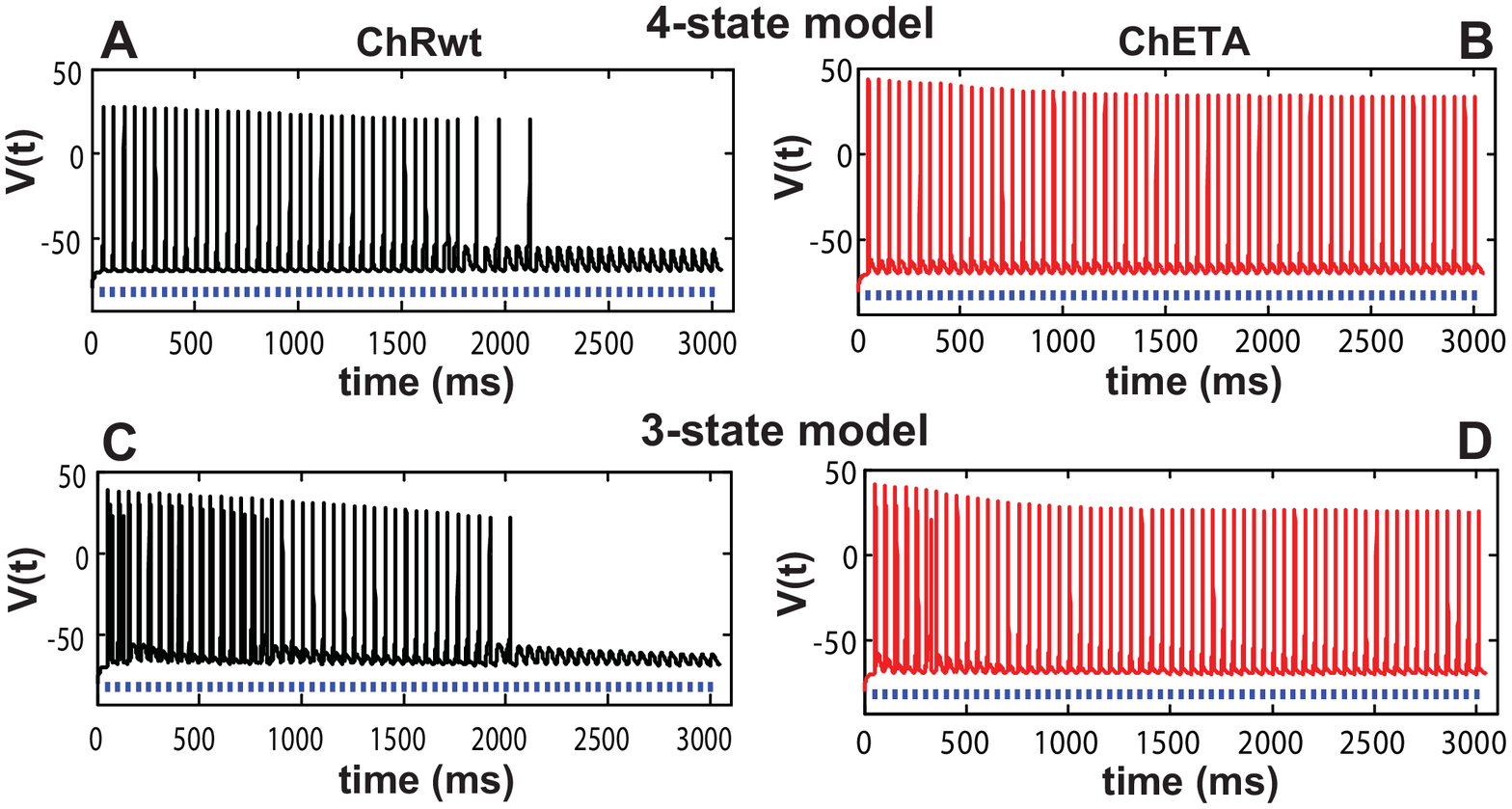}
  \caption{\textbf{Comparison of neural response to a prolonged optostimulation protocol.} The neural activity elicited in interneurons expressing ChRwt (black) and ChETA (red) by a train of 60 optostimuli presented at 20Hz, each pulse of 2 ms duration, is presented in comparison when the 3 and 4 state models are employed to account for ChR2 kinetics. \textbf{A} and \textbf{B} Response of interneurons expressing the two types of channelrhodopsins when the 4-state model is employed to mimic the ChR2 kinetics. The overall conductance was $g_{1} = 10.8$ (\textbf{A}) and $g_{1} = 70$ (\textbf{B}). These results are in good agrement with the experimental data presented din \citep{Gunaydin2010} Figure 3 e). \textbf{C} and \textbf{D} Similar results obtained when the 3-state model is used to account for ChR2 dynamics. The overall conductance was $g_{1}$ = 4 (\textbf{C}) respectively $g_{1}$ = 2.2 (\textbf{D}). }
  \label{Fig4}
\end{figure*}

We have also investigated the neural response of the WB model neuron to prolonged (more than 50 stimuli) optostimulation protocols. We observe that both the 3-state and the 4-state model can capture the behavior of spike failure observed in experimental data on interneurons expressing ChRwt (Figure \ref{Fig4} panels A and C in comparison with the experimental results presented in \citep{Gunaydin2010}, Figure 3e) and the correct neural response observed when the fast variant ChETA is expressed (see Figure \ref{Fig4}B and \ref{Fig4}D). However, with the 3-state model implementation, we observed a transient spike adaptation phenomenon in the spiking response of the WB model neuron at the beginning of the optostimulation protocol (more pronounced for 3-state model implementation of ChRwt), a phenomenon that has not been reported in experimental recordings \citep{Gunaydin2010}.

\begin{figure*}[htbp]
  \centering
    \includegraphics[scale=0.6]{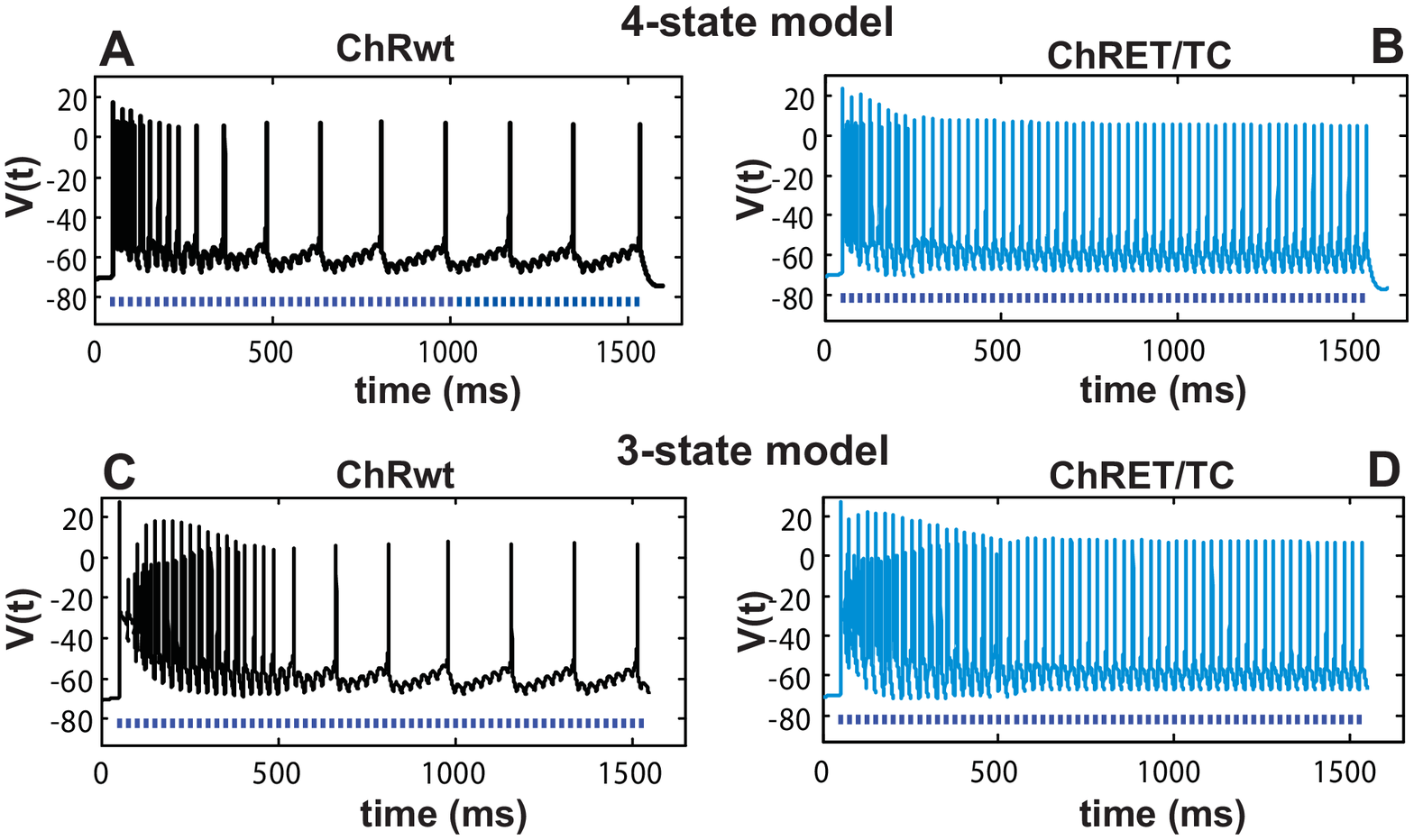}
  \caption{\textbf{Comparison of pyramidal response to a prolonged optostimulation protocol.} The simulated neural response to an optostimulation protocol comprising 60 stimuli of 2 ms each presented at 40 Hz is evaluate for different configurations. \textbf{A} and \textbf{B} Results obtained in pyramidal neuronal model by employing the 4-state model to account for ChRwt (\textbf{A}) and ChRET/TC (\textbf{B}) kinetics. The overall conductance was $g_{1} = 4.8$ (\textbf{A}) and $g_{1}=33$ (\textbf{B}). These results are consistent with the experimental findings reported in \citep{Berndt2011}, Figure 4 A. \textbf{C} and \textbf{D} Similar results obtained in pyramidal neurons when the 3-state model has been employed to account for ChR2 dynamics. The overall conductance was $g_{1} = 12.6$ (\textbf{C}) and $g_{1} = 12$ (\textbf{D}). }
  \label{Fig5}
\end{figure*}

We have also conducted a comparative study to evaluate the performance of the 3-state and 4-state models in mimicking experimental data on neural response of hippocampal pyramidal cells expressing ChRwt and a fast variant labeled ChRET/TC to various optostimulation protocols \citep{Berndt2011}. We incorporated a 3-state and 4-state model for ChRwt and the variant ChRET/TC into the Gol model for hippocampal pyramidal neurons \citep{Golomb2006}. We find that when the 3-state model for both ChRwt and ChRET/TC is implemented in the Gol model neuron, the model neuron elicits a significant number of additional spikes at the beginning of a 40 Hz optostimulation protocol, a feature that is inconsistent with experimental findings. On the other hand, the 4-state model implementation produces neural response that closely matches those observed experimentally (Figure \ref{Fig5}A and \ref{Fig5}B respectively in comparison to Figure 4A from \citep{Berndt2011}).

\begin{figure}[htbp]
  \centering
    \includegraphics[scale=0.5]{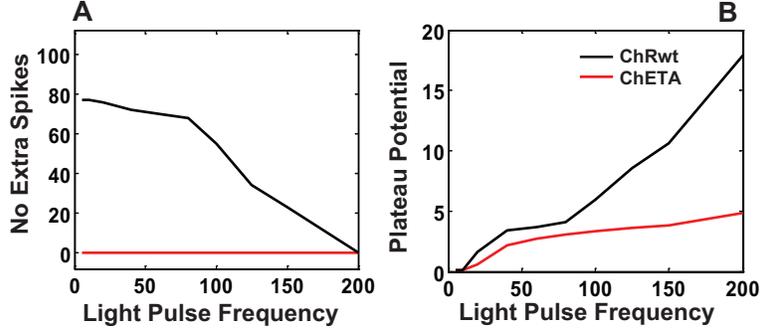}
  \caption{\textbf{Evaluation of the number of extra spikes and plateau potential.}\textbf{ A.} The number of extra spikes have been evaluated when optostimulation trains of 40 stimuli of 2 ms each and various frequencies have been applied on interneurons expressing ChRwt(black) and ChETA(red). In ChRwt simulations presented here $\tau_{ChR2} = 0.5$. The results are comparable with the experimental findings presented in \citep{Gunaydin2010} Figure 4b \textbf{ B.} Modeling results when the plateau potential is evaluated in the same conditions as in \textbf{A.} are qualitatively similar with the experimental results presented in \citep{Gunaydin2010} Figure 4c.}
  \label{Fig6}
\end{figure}

It is evident from the number of examples presented above that the 4-state model is better able to account for the multitude of neural response features elicited by optostimulation protocols in both excitatory and inhibitory model neurons. We therefore focus on the 4-state model and further investigate the extent to which several other features of neural activity generated by optostimulation can be captured using the 4-state model. We evaluate whether the number of extra spikes elicited by optostimulation protocols of different frequencies when 4- state models for ChRwt and ChETA variant are expressed in WB model interneuron match those reported in experiments of \citep{Gunaydin2010}. We find a good correspondence with the experimental findings for both the ChRwt and ChETA variants (Figure \ref{Fig6}A in comparison to Figure 4b in \citep{Gunaydin2010}). We also report on the ability of the 4-state models for ChRwt and ChETA to mimic experimental findings on the plateau potential elicited by the WB model neuron in response to optostimulation with various frequencies (Figure \ref{Fig6}B). As reported in the experimental findings of Gunaydin et al., (Figure 4c in \citep{Gunaydin2010}), we find that the WB model neuron expressing ChRwt consistently generates higher values for the plateau potential as compared to those generated by the WB model neuron expressing ChETA.

\begin{figure*}[htbp]
  \centering
    \includegraphics[scale=0.7]{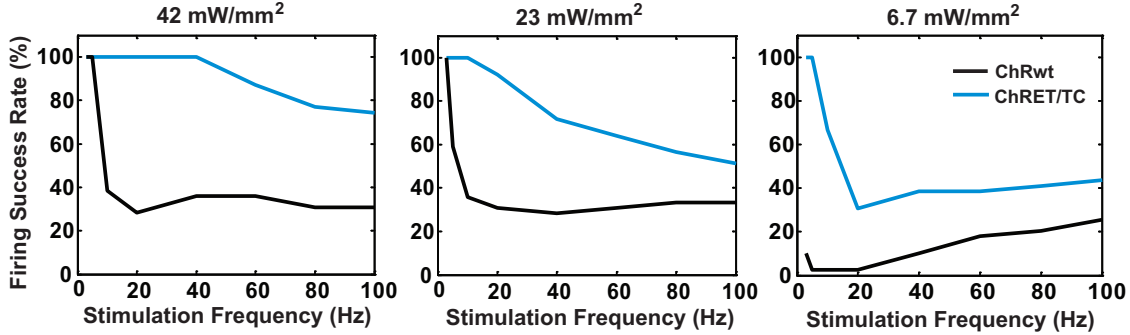}
  \caption{\textbf{Modeling results of neural response to different optostimulation frequencies with light of different intensities}. Modeling results of firing success rate simulated in pyramidal neurons expressing ChRwt (black) and ChRET/TC (blue) using the 4-state model and optostimulation protocols identical of 60 light pulses of 2 ms each and different stimulations frequencies. The results are comparable with the experimental findings presented in \citep{Berndt2011} Fig. 4B.}
  \label{Fig7}
\end{figure*}

Finally, we report results on the 4-state model's ability to capture the dependence of the firing success rate of a pyramidal neuron on the light intensity of optostimulation protocol (see Figure \ref{Fig7}). We find that the general trend, observed experimentally, of decrease in the success rate of optostimulation protocol to induce an action potential spike in pyramidal neuron is captured well in our simulation using the 4-state model for ChRwt and ChRET/TC incorporated in the Gol model neuron. However, we also find certain discrepancies. For example, for higher optostimulation frequencies ($>$60 Hz), for all light intensities tested, the firing success rate of the Gol model neuron using either of the two ChR2 variants is consistently higher than those reported in experiments (Figure 4B from \citep{Berndt2011} ). While, for low intensity (6.7 mW/mm$^{2}$) and low frequency ($<$ 30 Hz) optostimulation protocol, the firing success rate observed in Gol model neuron is lower than those reported in experiments. These inconsistencies may be due to processes such as the dependence of quantum efficiency of the ChR2 protein on light intensity ($\epsilon = \epsilon(I)$) or variability in the degree of absorbtion and diffraction as a function of light intensity intensity ($w_{loss} = w_{loss}(I)$), which are not captured by the 4-state model employed here.

\subsection{Degeneracy in the 4-state model}
Our analysis thus far suggests that the 4-state model for ChR2 photocurrent kinetics is better suited to account for the dynamical features of the experimentally observed neural response to various optostimulation protocols. This model gives rise to a general solution with three time constants in the light on condition and two time constants in the light off condition. We note however that the majority of experimental studies concerned with the molecular kinetics of light activated ion channels report a single time constant for the decay of the ChR2 photocurrent from peak to steady state when light is turned on and a single time constant for the decay of the photocurrent from steady state to zero when light is turned off. These data suggest that a 3-state model is an appropriate mathematical construct to mimic ChR2 photocurrent kinetics. However, as we have reported above, the 3-state model is unable to capture all the feature of neural response to optostimulation protocols. We also note that a most  generic 3-state model than considered here (not reported here) also fails to capture many of the key characteristics of neural response to optostimulation. The 4-state model can exhibit mono-exponential decay characteristics for the ChR2 photocurrent kinetics under the following three special conditions: (1) The solution to the 4-state model exhibits degenerate time constants and/or (2) the amplitude of the currents associated with one of the time constants is negligible and/or (3) there is at least an order of magnitude difference in the time constants associated with the solution of the 4-state model corresponding to a rapidly decaying current component, that is not captured in the empirical fit to the experimental data.

In order to evaluate whether any of these conditions are satisfied for the set of parameters reported in Table \ref{table3} for the 4-state model, we obtain an approximate analytical solution to equation \ref{eq9s} under the assumption that the conformational changes mediating the channel opening are considered instantaneous, i.e., $s=1$. This approximation holds as long as the time constant $\tau_{ChR2}$ is not significantly larger than the peak time of the photocurrent $t_{p}$, a condition satisfied by three of the ChR2 variants investigated in this paper (ChRwt and ChRET/TC modeled based on the data provided in \cite{Berndt2011} and ChETA modeled based on the data provided in \cite{Gunaydin2010}). We begin by identifying a semi-analytical solution for the \emph{light on} condition in the 4-state model (Appendix, Section 6.5 for details). The photocurrent can be expressed as:
\begin{eqnarray}
I_{\text{on}}(t) &=& I_{\text{rise(on)}}e^{-\Lambda_{\text{rise(on)}} t} + I_{\text{s(on)}}e^{-\Lambda_{\text{(on)}} t} \nonumber \\ &+& I_{\text{f(on)}}e^{-\Lambda_{\text{f(on)}} t} +I_{\text{plat}}
\end{eqnarray}
where, $\tau_{\text{rise(on)}}=\Lambda_{\text{rise(on)}}^{-1}$ represents the time constant associated with the rising phase of the photocurrent (also labeled as the activation time constant), $\tau_{\text{s(on)}}=\Lambda_{\text{s(on)}}^{-1}, \tau_{\text{f(on)}}=\Lambda_{\text{f(on)}}^{-1}$ are the slow and fast time constants associate with the bi-exponential photocurrent decay from peak to plateau (also referred to as the inactivation time constants), $I_{\text{rise(on)}}$ is the amplitude of the rising phase of the photocurrent, $I_{\text{s(on)}}$ is the amplitude of the slow component of the photocurrent decay from peak to plateau, $I_{\text{f(on)}}$ is the  amplitude of the fast component of the photocurrent decay from peak to plateau and $I_{\text{plat}}$ is the steady state plateau photocurrent. The expression for all these photocurrent components can be found in Section 6.4 of the Appendix.

For the \emph{light off} condition we use the solution provided in Nikolic et al. 2009 \citep{Nikolic2009} which reads:

\begin{equation}
I_{\text{off}} = I_{\text{s(off)}}e^{-\Lambda_{\text{s(off)}} t} + I_{\text{f(off)}}e^{-\Lambda_{\text{f(off)} t}};
\end{equation}

where $\tau_{\text{s(off)}}=\Lambda_{\text{s(off)}}^{-1}, \tau_{\text{f(off)}}=\Lambda_{\text{f(off)}}^{-1}$ are the slow and fast eigenvalues associate with the bi-exponential photocurrent decay from plateau to zero, $I_{\text{s(off)}}$ and $I_{\text{f(off)}}$ are the amplitudes of the slow and fast photocurrent decay from plateau to zero after the light is turned off. For both solutions the analytical expressions of the time constants and the current amplitudes can be found in Appendix, Section 6.5.

In Table \ref{table4}, we present the values of the fast and slow time constants and the corresponding amplitude values for the photocurrent decay from peak to plateau under the light on condition and from plateau to zero under the light off condition. We find that for each variant, for \emph{light off} condition, the  time decay of the slow component ($\tau_{s} = \Lambda_{s}^{-1}$) provided by the analytical solution is very similar in value with the one reported experimentally; the second (usually unreported) time decay ($\tau_{f} =\Lambda_{f}^{-1}$) is considerably smaller, implying that its associated photocurrent component decays on a much faster time scale relative to the experimental measurements. Furthermore, the amplitude of this photocurrent component is also significantly smaller as compared to  that of slower component of the photocurrent (see Table \ref{table4} for a comparison).

\begin{table*}
\begin{center}
     \begin{tabular}{|l|l|l|l|}
    \hline
              & \textbf{Light On }                           & \textbf{Light Off} \\ \hline
    ChRwt     & $\tau_{in} = 9.6$ \text{(experimental)}    & $\tau_{off} = 11.1$ \text{(experimental)} \\ \hline
              & $\tau_{s} = 10.91; A_{s} = 0.736$   & $\tau_{s} = 11.255; A_{s} = 0.036$ \\
              & $\tau_{f} = 7.47; A_{f} = -0.756$   & $\tau_{f} = 0.166; A_{f} = -0.0004$ \\ \hline
    ChETA    & $\tau_{in} = 15$  \text{(experimental)}    & $\tau_{off} = 5.2$ \text{(experimental)}\\ \hline
              & $\tau_{s} = 14.91; A_{s} = 0.0065$  & $\tau_{s} = 6.625; A_{s} = 0.0043$ \\
              & $\tau_{f} = 4.65; A_{f} = -0.0046$  & $\tau_{f} = 0.095; A_{f} = 0.0001$ \\ \hline
    ChRET/TC  & $\tau_{in} = 11$  \text{(experimental)}    & $\tau_{off} = 8.1$ \text{(experimental)}\\ \hline
              & $\tau_{s} = 8.11; A_{s} = 0.5457$   & $\tau_{s} = 8.36; A_{s} = 0.0105$ \\
              & $\tau_{f} = 7.17; A_{f} = -0.5494$  & $\tau_{f} = 0.06; A_{f} = -3.3e-005$ \\
    \hline
    \end{tabular}\vspace{0.2cm}
    \caption{\textbf{Comparison between the experimental and analytical time decay constants.} The photocurrent decay from peak to plateau $\tau_{in}$ and from plateau to zero ($\tau_{off}$) for each ChR2 variant is compared with the fast ($\tau_{f}$) and slow ($\tau{s}$ time constants)derived analytically. The maximum amplitude of the photocurrent component associated with each time decay is also provided.} \label{table4}
\end{center}
\end{table*}

For \emph{light on} condition ChRwt and ChRET/TC variant exhibit degeneracy, namely the fast and slow time constants are of same order in magnitude and their sum can be well-approximated using a single exponential decay function. For the variant ChETA no degeneracy is observed;  the slow time constant is very similar with the experimentally observed time decay while the fast photocurrent component has a significantly lower time constant and amplitude (conditions similar with the ones observed for \emph{light off} condition).

Together, these results provide clues to the absence of reports on the bi-exponential decay time constants for the ChR2 photocurrent profiles in majority of {\it in vitro} experiments and at the same time provide support for a 4-state model as an appropriate model to mimic ChR2 photocurrent kinetics of various types of light activated channels.

\section{Discussion}
In this paper we have systematically analyzed the 3-state and 4-state transition rate models for several ChR2 variants. Using experimental data available from recently published experimental studies we first identified the 3-state and 4-state model parameters.  We incorporate these models into model neurons and investigate the extent to which the dynamical features of model neuron response match to those observed experimentally for various optostimulation protocols. We demonstrate that the 3-state model presents significant limitations, not because of low model complexity as was reported earlier \citep{Nikolic2006}, but rather due to its inability to mimic the correct ratio $R$ of steady state to peak photocurrent in the presence of prolonged light stimulation and as a result its failure to account for
for certain dynamical features of neural activity elicited by optostimulation of hippocampal excitatory and inhibitory neurons. In comparison, the implementation of the 4-state model produced better match to the experimental data.

Two different strategies were adopted to estimate the parameters for 3-state and 4-state models from the 
available experimental data of ChR2 photocurrent profiles. The low complexity of 3-state model afforded analytical expressions to uniquely determine the 3-state model parameters in terms of experimental data. On the other hand, the increased complexity of 4-state model warranted empirical procedure to estimate all the model parameters. Motivated by the overall better performance of the 4-state model, we performed further investigations to determine the conditions under which a characteristic specific to the 3-state model, namely the mono-exponential decay of the ChR2 photocurrent kinetics (often reported in the experimental literature) can occur in the 4-state model.

For the variants investigated in this paper, there are two possible ways in which the 4-state model can account for the observed mono-exponential decay characteristics of ChR2 photocurrents: first, a degeneracy in the  \emph{light on} condition for two of the variants (ChRwt and ChRET/TC); second, or alternatively,  a significant difference in the two decay time constants with one decay time constant matching those reported experimentally while the second time constant as well as the associated amplitude component being small enough to remain unidentified in the empirical fit to the experimental photocurrent profile.  Together, these results suggest that the 4-state model is better suited to capture the complexity of the photocurrent kinetics of all ChR2 variants tested, possibly representing a generic mathematical framework to model the photocurrent dynamics of ChR2 proteins.

Our detailed computational analysis suggests that some neural dynamical features elicited by optostimulation depend on the degree of expression of the ChR2 variant in the cell. This dependence is more prominent for ChR2wt and can be observed for optostimulation protocols performed on interneurons  at rest (see Figure \ref{Fig3}). These observations suggest that in addition to the parameters accounting for the ChR2 kinetics (such as the activation and inactivation time constants,etc) the degree of ChR2 expression should be also considered a parameter of significant importance in computational models of neural activity elicited by optostimulation. It is also important to note that while employing the 4-state model provides results in excellent agreement with the experimental recordings across a multitude of optostimulation protocols, not all the features can be captured with a unique set of parameters. This aspect is particularly important for future optogenetic control strategies of neural function which generally rely on the consistency between the experimentally measured neural activity and the model based, predicted neural response.

Progress in the field of genetics and molecular biology has resulted in the identification of a myriad of light sensitive proteins with enhanced kinetic features and improved sensitivity to certain light wavelengths. For the purpose of this paper, we have chosen to discuss two fast variants (ChETA and ChRET/TC) in comparison with the more common ChRwt for the following reasons: 1) the experimental studies describing the response of these proteins to optostimulation provided all the necessary data to uniquely determine the parameters for the 3-state model and constrain the parametric search in the 4-state model; 2) these studies have also provided detailed information on the neural response (of two important classes of neurons, namely hippocampal pyramidal cells and PV-interneurons) to a multitude of optostimulation protocols, providing strong experimental datasets necessary for evaluating the efficiency of different model for ChR2 photocurrent kinetics. Evidently, the modeling methods outlined in this paper and the analysis thereof, can be easily implemented in the case of any other variant of potential interest in research applications.

Our results demonstrate the ability of 4-state transition rate model to mimic experimental findings across a multitude of optostimulation protocols applied on interneurons as well as pyramidal cells expressing different types of ChR2 variants. The ability to successfully bridge the experimental and computational observations is critical for future development of optostimulation control protocols. In this context, this paper, speaks in favor of the exciting possibility of using computational models to support and enhance the design of novel control strategies and optostimulation protocols to regulate normal and abnormal neural network activity.

\section{Acknowledgements}

We would like to thank Lisa Gunaydin and Andre Berndt for sharing their data with us. This research was funded by startup funds to SST; The intramural grant on Computational Biology at the University of Florida; and the Wilder Center of Excellence for Epilepsy Research and the Children's Miracle Network. PPK was partially supported by the Eckis Professor Endowment at the University of Florida.

\section{Appendix}

\subsection{Analytical solution of the 3-state model.}

The equations describing the model are:

\begin{eqnarray}\label{eq1a}
\dot{o} &=& P(1-o-d) - G_{d}o\\
\dot{d} &=& G_{d}o - G_{r}d\nonumber
\end{eqnarray}

The photocurrent is:

\begin{equation}\label{eq2a}
I = Vg_{1}o
\end{equation}

The equivalent theoretical solution is:

\begin{equation}\label{eq3a}
I = C_{1}e^{-\lambda_{1}t} + C_{2}e^{-\lambda_{2}t} + I_{plat}
\end{equation}

where

\begin{equation}\label{eq4a}
\lambda_{1} = \alpha - \beta; \qquad \lambda_{2} = \alpha + \beta
\end{equation}

\begin{equation}\label{eq5a}
\alpha = \frac{P+G_{d}+G_{r}}{2}; \qquad \beta = \sqrt{\frac{(P-G_{d}-G_{r})^{2}}{2} - G_{d}G_{r}}; \qquad I_{plat} = \frac{P(\lambda_{1} + \lambda_{2} - P - G_{d})}{\lambda_{1}\lambda_{2}}\\
\end{equation}

and

\begin{eqnarray}\label{eq6a}
C_{1} = \frac{D_{0}P\lambda_{1} + (\lambda_{2} -P -G_{d})(P - O_{0}\lambda_{1})}{\lambda_{1}(\lambda_{1}-\lambda_{2})}\\
C_{2}=O_{0} - C_{1} + \frac{P(P+G_{d}-\lambda_{1}-\lambda_{2})}{\lambda_{1}\lambda_{2}}\nonumber
\end{eqnarray}

When ideal initial conditions are satisfied ($D_{0} = 0; O_{0} = 0$) the solution presented in \cite{Nikolic2009} is recovered.

\subsection{Derivation of formula (\ref{eq8s}) from the main manuscript.}

We start with the expression of the first eigenvalue given by equation \ref{eq4} above:

\begin{equation}\label{eq7a}
\lambda_{1} = \frac{P+G_{d}+G_{r}}{2} - \frac{\sqrt{(P-G_{d}-G_{r})^{2} - 4G_{d}G_{r}}}{2}
\end{equation}

We multiply the equation above with 2 and rearrange the terms to obtain:

\begin{equation}\label{eq8a}
P+G{d}+G_{r} - 2\lambda_{1} = \sqrt{(P-G_{d}-G_{r})^{2} - 4G_{d}G_{r}}
\end{equation}

we raise the equation above to the second power and rearrange the terms to obtain:

\begin{equation}\label{eq9a}
\lambda_{1}^{2} - \lambda_{1}(G_{d}+G_{r}) +P(G_{d}+G_{r} - \lambda_{1}) +G_{d}G_{r} = 0;
\end{equation}

which finally gives:

\begin{equation}\label{eq10a}
P = \lambda_{1} + \frac{G_{r}G_{d}}{\lambda_{1} - G_{r} - G_{d}}
\end{equation}

\subsection{Evaluation of necessary initial conditions for the 3-state model.}

We can find the initial conditions necessary to obtain a photocurrent which will exhibit the appropriate $I_{peak}/I_{plat}$ ratio by use of the following conditions:\\

1. We first find the coefficients of the homogeneous solution $(C_{1}, C_{2})$ which will satisfy the experimental data by solving the following system of equations:

\begin{eqnarray}\label{eq11a}
\lambda_{1}C_{1}e^{-\lambda_{1}t_{p}} + \lambda_{2}C_{2}e^{-\lambda_{2}t_{p}} = 0\\
\frac{C_{1}e^{-\lambda_{1}t_{p}}}{I_{plat}} + \frac{C_{2}e^{-\lambda_{2}t_{p}}}{I_{plat}} + 1 = \frac{1}{R}\nonumber
\end{eqnarray}

where the first equation represents the condition that the derivative of the photocurrent function is zero for the $I = I_{peak}$ and the second equation instantiate the condition that the ration $I_{peak}/I_{plat}$ must match the ratio $\frac{1}{R}$ provided by the experimental data. \\

2. With $C_{1}$ and $C_{2}$ determined above we can find the initial conditions $(o_{0}, o_{0})$ by solving the equations 7 and 8 given in the previous section; then $C_{0} = 1-o_{0} - d_{0}$.\\

The solutions for the all of the above equations have been evaluated symbolically in Matlab by using the function \textbf{solve} and then numerically by allowing the parameters to take the appropriate values in the symbolic solution.

\subsection{Experimental Results - Additional Information.}
\subsubsection{Evaluation of the activation time constant $\tau_{rise}$.}
The evaluation of the time constant of the rising phase of the photocurrent from zero to peak has been performed by approximating the photocurrent curve with a mono-exponential function. Thus, we can write:

\begin{equation}\label{eq12a}
I(t)  = I_{max}(1-e^{-\frac{t}{\tau_{rise}}})
\end{equation}

when the the photocurrent reaches maximum (at $t = t_{p}$) we can approximate:

\begin{equation}\label{eq13a}
\frac{I}{I_{max}}\simeq 0.99999;
\end{equation}

which leads to:

\begin{equation}\label{eq14a}
\tau_{rise} = -\frac{t_{p}}{ln(0.00001)}
\end{equation}

\subsubsection{Comparison between the photocurrent induced by continuous 1s and brief 2 ms optostimulation in cell expressing ChRwt and the fast ChRET/TC variant.}

\begin{figure*}[htbp]
  \centering
    \includegraphics[scale=0.7]{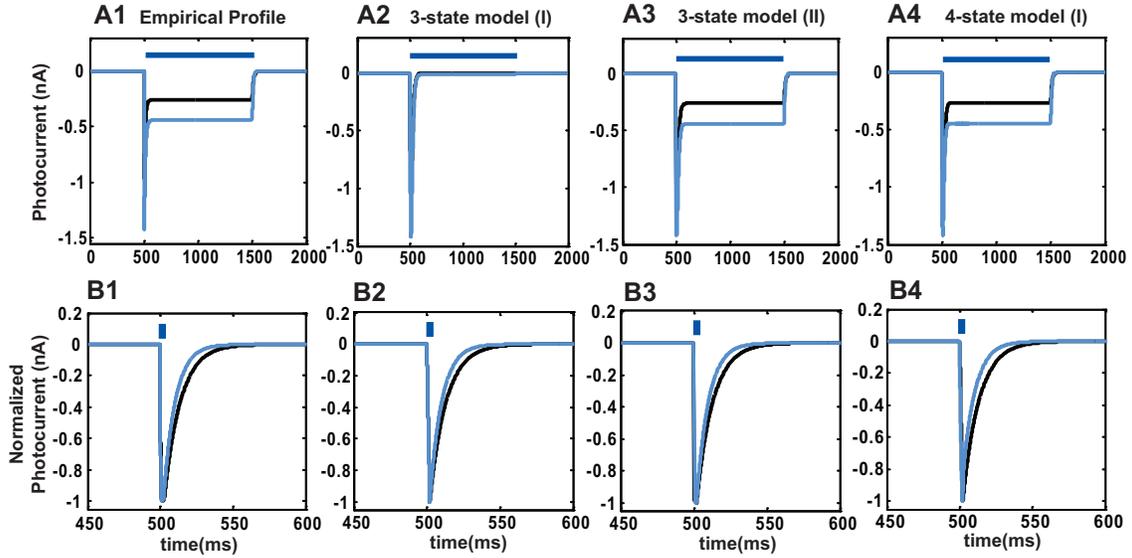}
  \caption{\textbf{Comparison example of the ChR2 photocurrent elicited by 1s and 2 ms optostimulation}. \textbf{A1} and \textbf{B1}. Empirical data profile constructed for ChRwt and ChRET/TC variant using equations 11 and 12 from the main paper as well as the experimental data provided in Table1 is displayed for a continuous 1s optostimulation (\textbf{A1}) and for a brief 2 ms optostimulation (\textbf{B1}). The two normalized experimental photocurrent profiles are matching the experimental results reported in \cite{Gunaydin2010} Fig 3c and Fig 2d. \textbf{A2} and \textbf{B2}. Photocurrent generated by the 3-state model starting from ideal initial conditions ((I): $C(0) = 1; O(0) = 0; D(0) = 0.$ ) for ChRwt and ChRET/TC variant for 1s (\textbf{A2}) respectively 2ms (\textbf{B2}) optostimulation. \textbf{A3} and \textbf{B3}. Photocurrent generated for ChRwt in comparison with ChR ET/TC, by the same 3-state model but starting from special initial conditions ((II): ChRwt: $C(0) = 0.0041; O(0) = 0.0037; D(0) = 0.9922$; ChRET/TC: $C(0) = 0.00156; O(0) = 0.0098; D(0) = 0.9746.$) evaluated in Appendix, Section 6.3. \textbf{A4} and \textbf{B4} Photocurrent generated for the same ChR2 variants using the 4-state model with ideal initial conditions ((I): $C_{1}(0) = 1; O_{1}(0) = 0; O_{2} = 0.; C_{2}(0) = 0.$ ) ).}
  \label{fig1}
\end{figure*}

We present in Fig.\ref{fig1} a comparison between the ChR wt and ChRET/TC photocurrent elicited by 1s and 2 ms continuous optostimulation.

\subsubsection{Dependence between the excitation rate (P) and light intensity (I).}

\begin{equation}\label{eq15a}
P = \epsilon F = \epsilon \frac{\sigma_{ret}\phi}{w_{loss}} = \frac{\epsilon \sigma_{ret}\lambda}{w_{loss}hc}I
\end{equation}

where $\epsilon$ is the quantum efficiency of photon absorbtion (a typical value for rhodopsin is $\epsilon \simeq 0.5$ \cite{Braun1999}), $F =\sigma_{ret}\phi/w_{loss}$ is the number of photons absorbed by ChR2 molecule per unit time, $\sigma_{ret}$ is the retinal cross-section ($\sigma_{ret}\simeq 1.2\times10^{-20}m^{2}$ \cite{Heg2005}), $w_{loss}$ is the measure of the loss of incidental photons due to scattering and absorbtion phenomena, $\phi = \lambda I/hc$ is the photon flux per unit area, $\lambda \simeq 480 nm$ is the wave length of the light used in the stimulation protocol, $I(mW/mm^{2})$ is the light intensity, $h = 6.626\times10^{-34}Js$ is the Planck's constant and $c = 3\times10^{8}m/s$ is the speed of light in vacuum.

\subsection{Derivation of the semi-analytical solution for the light on condition in the 4-state model.}

The 4 state model can be written as follows:
\begin{eqnarray}\label{eq16a}
\dot{o_{1}} &=& P_{1}(1-c_{2}-o_{1}-o_{2}) - (G_{d1}+e_{12})o_{1} + e_{21}o_{2}\nonumber\\
\dot{o_{2}} &=& P_{2}c_{2} + e_{12}o_{1} - (G_{d2}+e_{21})o_{2}\\
\dot{c_{2}} &=& G_{d2}o_{2} - (P_{2}+G_{r})c_{2}\nonumber
\end{eqnarray}

With the notation $y = [o_{1}\quad o_{2}\quad c_{2}]^{T}$, equation \ref{eq16a} can be than expressed as follows:

\begin{eqnarray}\label{eq18a}
\dot{y} &=&
\left[
\begin{matrix}
    -(P_{1}+G_{d1}+e_{12})& e_{21}-P_{1} & -P_{1} \\
    e_{12} & -(G_{d2}+e_{21}) & P_{2} \\
    0 & G_{d2} & -(P_{2}+G_{r})
\end{matrix} \right]y +
\left[
\begin{matrix}
P_{1}\\
0\\
0
\end{matrix}\right]\\ \nonumber \\ \nonumber
%
&=& Ay+P  \nonumber
\end{eqnarray}

The general solution that needs to be evaluated can be written as:

\begin{equation}\label{eq19a}
y = y_{c}+y_{p}
\end{equation}

where $y_{c}$ of the homogeneous (complementary) solution and $y_{p}$ is the particular solution of the system of equations (18). In the following we will evaluate both components.\\

\subsubsection{Finding the eigenvalues.}

The characteristic equation is:

\begin{equation}\label{eq20a}
det(A - \lambda I) = 0
\end{equation}
or
\begin{equation}\label{eq21a}
\left|
\begin{matrix}
    -(P_{1}+G_{d1}+e_{12}) - \lambda & e_{21}-P_{1} & -P_{1} \\
    e_{12} & -(G_{d2}+e_{21}) - \lambda & P_{2} \\
    0 & G_{d2} & -(P_{2}+G_{r}) - \lambda
\end{matrix} \right| =0
\end{equation}
which leads to
\begin{eqnarray}\label{eq22a}
-[(P_{1}+G_{d1}+e_{12}+\lambda)(G_{d2}+e_{21}+\lambda)] - P_{1}e_{12}G_{d2} + \\ P_{2}G_{d2}(P_{1}+G_{d1}+e_{12}+\lambda) + e_{12}(e_{21}-P_{1})(P_{2}+G_{r}+\lambda) = 0 \nonumber
\end{eqnarray}

This equation is solved symbolically in Matlab using the commend \textbf{solve} which gives the expressions for the solutions: $\lambda_{1}, \lambda_{2}, \lambda_{3}$. The actual expressions are very elaborated, therefore they will not be included here. The numerical evaluation of these eigenvalues has been performed in Matlab using the function \textbf{eval} and the parameter values provided for each variant in the main paper, Table 3.\\

\subsubsection{Finding the eigenvectors.}

The characteristic equation is:

\begin{equation}\label{eq23a}
Av = \lambda v
\end{equation}
or
\begin{equation}\label{eq24a}
\left[
\begin{matrix}
    -(P_{1}+G_{d1}+e_{12})& e_{21}-P_{1} & -P_{1} \\
    e_{12} & -(G_{d2}+e_{21}) & P_{2} \\
    0 & G_{d2} & -(P_{2}+G_{r})
\end{matrix} \right]
\left[
\begin{matrix}
v_{1}\\
v_{2}\\
v_{3}
\end{matrix}\right] =
\lambda_{1}\left[
\begin{matrix}
v_{1}\\
v_{2}\\
v_{3}
\end{matrix}\right]
\end{equation}

Then the eigenvectors satisfying this equation are:
\begin{equation}\label{eq25a}
v =
\left[
\begin{matrix}
1\\
\frac{(P_{2}+G_{r}+\lambda_{1})(\lambda_{1}+P_{1}+G_{d1}+e_{12})}{(e_{21}-P_{1})(P_{2}-G_{r}+\lambda_{1})-P_{1}G_{d2}}\\
\frac{G_{d2}(\lambda_{1}+P_{1}+G_{d1}+e_{12})}{(e_{21}-P_{1})(P_{2}-G_{r}+\lambda_{1})-P_{1}G_{d2}}
\end{matrix}\right];
%
%
u=
\left[
\begin{matrix}
1\\
\frac{(P_{2}+G_{r}+\lambda_{2})(\lambda_{2}+P_{1}+G_{d1}+e_{12})}{(e_{21}-P_{1})(P_{2}-G_{r}+\lambda_{2})-P_{1}G_{d2}}\\
\frac{G_{d2}(\lambda_{2}+P_{1}+G_{d1}+e_{12})}{(e_{21}-P_{1})(P_{2}-G_{r}+\lambda_{2})-P_{1}G_{d2}}
\end{matrix}\right];
%
%
w=
\left[
\begin{matrix}
1\\
\frac{(P_{2}+G_{r}+\lambda_{3})(\lambda_{3}+P_{1}+G_{d1}+e_{12})}{(e_{21}-P_{1})(P_{2}-G_{r}+\lambda_{3})-P_{1}G_{d2}}\\
\frac{G_{d2}(\lambda_{3}+P_{1}+G_{d1}+e_{12})}{(e_{21}-P_{1})(P_{2}-G_{r}+\lambda_{3})-P_{1}G_{d2}}
\end{matrix}\right].
\end{equation}

\subsubsection{The complementary solution.}

The complementary solution can then be written as:

\begin{equation}\label{eq26}
y_{c} = C_{1}e^{\lambda_{1}t}v + C_{2}e^{\lambda_{2}t}u + C_{3}e^{\lambda_{3}t}w
\end{equation}
or
\begin{equation}\label{eq27}
y_{c} =
\left[
\begin{matrix}
e^{\lambda_{1}t}&e^{\lambda_{2}t}&e^{\lambda_{3}t}\\
\frac{(P_{2}+G_{r}+\lambda_{1})(\lambda_{1}+P_{1}+G_{d1}+e_{12})}{(e_{21}-P_{1})(P_{2}-G_{r}+\lambda_{1})-P_{1}G_{d2}}e^{\lambda_{1}t}&
\frac{(P_{2}+G_{r}+\lambda_{2})(\lambda_{2}+P_{1}+G_{d1}+e_{12})}{(e_{21}-P_{1})(P_{2}-G_{r}+\lambda_{2})-P_{1}G_{d2}}e^{\lambda_{2}t}&
\frac{(P_{2}+G_{r}+\lambda_{3})(\lambda_{3}+P_{1}+G_{d1}+e_{12})}{(e_{21}-P_{1})(P_{2}-G_{r}+\lambda_{3})-P_{1}G_{d2}}e^{\lambda_{3}t}\\
\frac{G_{d2}(\lambda_{1}+P_{1}+G_{d1}+e_{12})}{(e_{21}-P_{1})(P_{2}-G_{r}+\lambda_{1})-P_{1}G_{d2}}e^{\lambda_{1}t}&
\frac{G_{d2}(\lambda_{2}+P_{1}+G_{d1}+e_{12})}{(e_{21}-P_{1})(P_{2}-G_{r}+\lambda_{2})-P_{1}G_{d2}}e^{\lambda_{2}t}&
\frac{G_{d2}(\lambda_{3}+P_{1}+G_{d1}+e_{12})}{(e_{21}-P_{1})(P_{2}-G_{r}+\lambda_{3})-P_{1}G_{d2}}e^{\lambda_{3}t}
\end{matrix}\right]\left[
\begin{matrix}
C_{1}\\
C_{2}\\
C_{3}
\end{matrix}\right]
\end{equation}
or, by notation:
\begin{equation}\label{eq28}
y_{c} = A_{c}C.
\end{equation}

\subsubsection{Finding the particular solution}

The following \emph{\textbf{analytical}} steps are performed in Matlab:

a) We evaluate the inverse of the complementary matrix $A_{c}^{-1}$;\\

b) We evaluate the product:

\begin{equation}\label{eq29}
Z = A_{c}^{-1}*P = A_{c}^{-1}\left[
\begin{matrix}
P_{1}\\
0\\
0
\end{matrix}\right]
\end{equation}

c) We integrate symbolically Z using the commend: $R = int(Z,t)$;\\

d) We evaluate the particular solution: $y_{p} = A_{c}R$.\\

\subsubsection{Finding coefficients $C_{1},C_{2},C_{3}$ by use of initial conditions.}

To coefficients of the homogeneous solution can be found by considering the following condition:

\begin{equation}\label{eq30}
A_{c}(0)C +y_{p}(0) = y_{0};
\end{equation}

The above system of equations cannot be solved in Matlab straightforward as the function \textbf{solve} is unable to find a solution when all three equations are presented simultaneously; the system can be solved however, if the analytical expression of two of the coefficients is given and only one equation is solved symbolically. For this purpose, we derive the equations for the first and second coefficient and solve symbolically in Matlab only for the last coefficient. We expand the above equation as follows:

\begin{equation}\label{eq31}
\left[
\begin{matrix}
A_{c_{11}}(0)&A_{c_{12}}(0)&A_{c_{13}}(0)\\
A_{c_{21}}(0)&A_{c_{22}}(0)&A_{c_{23}}(0)\\
A_{c_{31}}(0)&A_{c_{32}}(0)&A_{c_{33}}(0)
\end{matrix}\right]\left[
\begin{matrix}
C_{1}\\
C_{2}\\
C_{3}
\end{matrix}\right] + \left[
\begin{matrix}
y_{p_{1}}(0)\\
y_{p_{2}}(0)\\
y_{p_{3}}(0)
\end{matrix}\right] = \left[
\begin{matrix}
0\\
0\\
0
\end{matrix}\right]
\end{equation}
or:
\begin{eqnarray}\label{eq32}
A_{c_{11}}(0)C_{1} + A_{c_{12}}(0)C_{2} + A_{c_{13}}(0)C_{3} +y_{p_{1}}(0) = 0\\\nonumber
A_{c_{21}}(0)C_{1} + A_{c_{22}}(0)C_{2} + A_{c_{23}}(0)C_{3} +y_{p_{2}}(0) = 0\\\nonumber
A_{c_{31}}(0)C_{1} + A_{c_{32}}(0)C_{2} + A_{c_{33}}(0)C_{3} +y_{p_{3}}(0) = 0
\end{eqnarray}

We get:

\begin{equation}\label{eq33}
C_{1} = (-y_{p_{1}}(0) - A_{c_{12}}(0)C_{2} - A_{c_{13}}(0)C_{3})/A_{c_{11}}(0);
\end{equation}

and

\begin{equation}\label{eq34}
C_{2} = \frac{A_{c_{21}}(0)y_{p_{1}}(0) - A_{c_{11}}(0)y_{p_{2}}(0) - C_{3}(A_{c_{23}}(0)A_{c_{11}}(0) - A_{c_{21}}(0)A_{c_{13}})(0)}{A_{c_{22}}(0)A_{c_{11}}(0) - A_{c_{21}}(0)A_{c_{12}}(0)}
\end{equation}

We introduce equations \ref{eq33} and \ref{eq34} in the last equation \ref{eq32} and then solve symbolically this equation in Matlab for $C_{3}$; we then evaluate $C_{1}$ and $C_{2}$.\\

We are now able to evaluate the full theoretical solution:
\begin{eqnarray}\label{eq35}
o_{1} &=& C_{1}e^{\lambda_{1}t} +C_{2}e^{\lambda_{2}t}+C_{3}e^{\lambda_{3}t}+y_{p_{1}}\\
o_{2} &=& A_{c_{21}}(0)C_{1}e^{\lambda_{1}t} + A_{c_{22}}(0)C_{2}e^{\lambda_{2}t}+A_{c_{23}}(0)C_{3}e^{\lambda_{3}t} + y_{p_{2}};\nonumber
\end{eqnarray}

The photocurrent elicited in light on condition will be:

\begin{equation}\label{eq36}
I = Vg_{1}(o_{1}+\gamma o_{2}).
\end{equation}
or
\begin{equation}\label{eq37}
I = Vg_{1}[C_{1}(1+\gamma A_{c_{21}}(0))e^{\lambda_{1}t} + C_{2}(1+\gamma A_{c_{22}}(0))e^{\lambda_{2}t} + C_{3}(1+\gamma A_{c_{23}}(0))e^{\lambda_{3}t}] + Vg_{1}(y_{p1}+\gamma y_{p2})
\end{equation}

The reader should note that all the eigenvalues above are negative. Therefore we can write:

\begin{equation}\label{eq38}
\lambda_{1} = -\Lambda_{1};\qquad \lambda_{2} = -\Lambda_{2}; \qquad \lambda_{3} = -\Lambda_{3};
\end{equation}

where $\Lambda_{1},\Lambda_{2},\Lambda_{3}>0$.

We evaluate the time constants associated with these eigenvalues to be:

\begin{equation}\label{eq39}
\tau_{1} = \frac{1}{\Lambda_{1}}; \qquad \tau_{2} = \frac{1}{\Lambda_{2}}; \qquad \tau_{3} = \frac{1}{\Lambda_{3}}
\end{equation}

The smallest time constant is associated with the rise phase (from zero to peak) of the photocurrent (the the beginning of the optostimulation); the other two controls the fast and slow component of the photocurrent decay from peak to steady state. With the parameters found in the manuscript Table 3 we identify:

\begin{equation}\label{eq40}
\tau_{rise(on)} = \tau_{2};\qquad \tau_{s(on)} = \tau_{1};\qquad \tau_{f(on)} = \tau_{3}
\end{equation}

\begin{equation}\label{eq41}
A_{rise(on)} = C_{2}(1+\gamma A_{c_{22}}(0)); \qquad A_{s(on)} = C_{1}(1+\gamma A_{c_{21}}(0)); \qquad A_{f(on)} = C_{3}(1+\gamma A_{c_{23}}(0))
\end{equation}

\begin{equation}\label{eq42}
I_{rise(on)} = Vg_{1}A_{rise(on)}; \qquad I_{s(on)} = Vg_{1}A_{s(on)};\qquad I_{f(on)} = Vg_{1}A_{f(on)}; \qquad I_{plat} = Vg_{1}(y_{p1} + \gamma y_{p2})
\end{equation}

\begin{equation}\label{eq43}
I_{on}(t) = I_{rise(on)}e^{-\Lambda_{rise(on)} t} + I_{s(on)}e^{-\Lambda_{s(on)} t} + I_{f(on)}e^{-\Lambda_{f(on)} t} +I_{plat};
\end{equation}

For convenience we reproduce here the analytical solution for light off condition provided in \cite{Nikolic2009}:

\begin{equation}\label{eq44}
\Lambda_{1} = b-c; \qquad \Lambda_{2} = b+c;
\end{equation}

\begin{equation}\label{eq45}
b = \frac{G_{d1}+G_{d2}+e_{12}+e_{21}}{2}; \qquad c = \sqrt{b^{2}-(G_{d1}G_{d2}+G_{d1}e_{21}+G_{d2}e_{12})}
\end{equation}

\begin{equation}\label{eq46}
I_{off} = I_{s(off)}e^{-\Lambda_{s(off)} t} + I_{f(off)}e^{-\Lambda_{f(off)} t};
\end{equation}

where

\begin{equation}\label{eq47}
I_{s(off)} = Vg_{1}A_{s(off)};\qquad I_{f(off)} = Vg_{1}A_{f(off)};
\end{equation}

and

\begin{eqnarray}\label{eq48}
A_{s(off)} = \frac{[\Lambda_{2} - (G_{d1}+(1-\gamma)e_{12})]O_{10} + [(1-\gamma)e_{21} + \gamma(\Lambda_{2}-G_{d2})]O_{20}}{\Lambda_{2}-\Lambda_{1}}\\
A_{f(off)} = \frac{[G_{d1} + (1-\gamma)e_{12} - \Lambda_{1}]O_{10} + [\gamma(G_{d2}-\Lambda_{1}) - (1-\gamma)e_{21}]O_{20}}{\Lambda_{2}-\Lambda_{1}}\nonumber
\end{eqnarray}

\newpage

\end{document}